\documentclass[]{fairmeta}

\usepackage{amsmath,amsfonts,bm}

\def\eqref#1{equation~\ref{#1}}

\def\1{\bm{1}}

\def\vz{{\bm{z}}}

\DeclareMathAlphabet{\mathsfit}{\encodingdefault}{\sfdefault}{m}{sl}
\SetMathAlphabet{\mathsfit}{bold}{\encodingdefault}{\sfdefault}{bx}{n}

\newcommand{\ie}{\textit{i}.\textit{e}., }
\newcommand{\eg}{\textit{e}.\textit{g}.\ }
\usepackage{float}
\usepackage{graphicx}
\graphicspath{{figs/}}
\usepackage{multirow}
\usepackage[utf8]{inputenc} 
\usepackage[T1]{fontenc}    
\usepackage{hyperref}       
\usepackage{url}            
\usepackage{booktabs}       
\usepackage{amsfonts}       
\usepackage{nicefrac}       
\usepackage{microtype}      
\usepackage{xcolor}         
\usepackage{tablefootnote} 

\definecolor{eeg}{RGB}{51,111,162}  
\definecolor{meg}{RGB}{58,146,58}  
\definecolor{fmri3t}{RGB}{180,46,50}  
\definecolor{fmri7t}{RGB}{223,193,32}  

\title{Scaling laws for decoding images from brain activity}

\author[1,*]{Hubert Banville}
\author[1,*]{Yohann Benchetrit}
\author[1]{Stéphane d'Ascoli}
\author[1]{Jérémy Rapin}
\author[1]{Jean-Rémi King}

\affiliation[1]{Meta AI}

\contribution[*]{Equal contribution.}

\date{\today}
\correspondence{\email{\{hubertjb,ybenchetrit,sdascoli,jrapin,jeanremi\}@meta.com}}

\abstract{
Generative AI has recently propelled the decoding of images from brain activity.
How do these approaches scale with the amount and type of neural recordings?
Here, we systematically compare image decoding from four types of non-invasive devices: electroencephalography (EEG), magnetoencephalography (MEG), high-field functional Magnetic Resonance Imaging (3T fMRI) and ultra-high field (7T) fMRI. 
For this, we evaluate decoding models on the largest benchmark to date, encompassing 8 public datasets, 84 volunteers, 498 hours of brain recording and 2.3 million brain responses to natural images. 
Unlike previous work, we focus on single-trial decoding performance to simulate real-time settings.
This systematic comparison reveals three main findings.
First, the most precise neuroimaging devices tend to yield the best decoding performances, when the size of the training sets are similar.
However, the \emph{gain} enabled by deep learning -- in comparison to linear models -- is obtained with the noisiest devices.
Second, we do not observe any plateau of decoding performance as the amount of training data increases. Rather, decoding performance scales log-linearly with the amount of brain recording.
Third, this scaling law primarily depends on the amount of data per subject. However, little decoding gain is observed by increasing the number of subjects. 
Overall, these findings delineate the path most suitable to scale the decoding of images from non-invasive brain recordings. 
}

\begin{document}

\maketitle

\section{Introduction}

\begin{figure}
    \centering
    \includegraphics[width=1.0\linewidth]{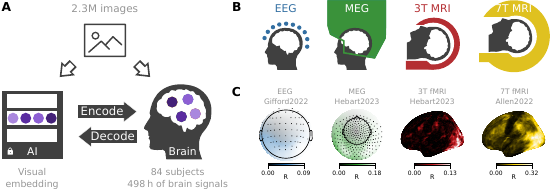}
    \caption{(\textbf{A}) Brain-to-image decoding and encoding pipeline. In decoding, brain models are trained to predict, from brain activity, the embeddings of the images learned by a pretrained computer vision model. Decoding predictions can then be fed to an image generation model to reconstruct the images. 
    In encoding, models are instead trained to predict brain activity from image embeddings.
    (\textbf{B}) Our analyses rely on multiple datasets of brain data and image pairs, focusing on four neuroimaging devices: EEG, MEG, 3T fMRI and 7T fMRI.
    (\textbf{C}) We validate the content of the datasets using encoding models trained to predict each M/EEG channel or fMRI voxel from the presented images, which yield the expected spatial response over the occipital region as measured with Pearson correlation. See \Cref{sec:encoding} for more details.
    }
    \label{fig:pipeline}
\end{figure}

Decoding natural images from brain activity originated in the 2000s \citep{kamitani2005decoding,miyawaki2008visual,naselaris2009bayesian} but progressed rapidly over the past two years.
Reconstructing images from functional Magnetic Resonance Imaging (fMRI) \citep{ozcelik2023brain,mai2023unibrain,zeng2023controllable,ferrante2022semantic, scotti2024mindeye2} and magnetoencephalography (MEG) \citep{benchetrit2024brain} 
can now be achieved by training a deep neural network to predict, from brain signals, the latent representation of an image, and then using this prediction to condition an image generation model (see \Cref{fig:pipeline}A).

Four main factors are presumably responsible for this recent progress. First, recent studies train their decoders on a larger amount of data than in the past: it is now common to train models on several hours of brain recordings per subject~\citep{defossez2022decoding, gwilliams2023introducing, schoffelen2019subject, tang2023semantic, armeni2022hour, allen2022massive, scotti2024mindeye2, benchetrit2024brain, hebart2023things, chehab2022deep}. Second, several approaches benefited from hardware improvements, such as the development of ultra high field (7T) fMRI \citep{allen2022massive}. 
Third, modern deep learning has effectively provided neuroscience with powerful representations of images in the brain \citep{kriegeskorte2016inferring, yamins2014performance, eickenberg2017seeing, schrimpf2018brain}. Indeed, computer vision models like OpenAI CLIP~\citep{radford2021learning, ozcelik2023brain, conwell2022can} and DINOv2 \citep{oquab2023dinov2, benchetrit2024brain, adeli2023predicting}, have been shown to learn representations that linearly predict brain responses to natural images. Fourth, recent models based on diffusion effectively help reconstructing plausible images from the decoding of these latent image features.

What are the best paths to improve brain-to-image decoding? In spite of a flourishing field, this issue is particularly difficult to address. 
First, most studies focus on a single neuroimaging device, \ie either electroencephalography (EEG), MEG, fMRI at 3~Tesla (3T fMRI) or 7T fMRI.
Second, existing datasets contain different numbers of subjects and of recordings per subject.
Third, each study uses different preprocessing steps, some of which are incompatible with single-trial (and therefore, real-time) evaluation of decoding pipelines.
Fourth, most studies use different image generation models -- making it difficult to credit any improvements to the data, the original method, or, more trivially, to a better image generation model.
Finally, the evaluation metrics are often disparate. In sum, this methodological variability obscures the factors critical for brain-to-image decoding. 

Here, we address this issue by disentangling the specific contributions of neuroimaging devices, data quantity and pretrained models. To achieve this, we systematically compare the decoding performance obtained using a variety of experimental setups within a unified benchmark. Our analysis is based on the brain activity of 84 healthy volunteers who watched a total of 2.3M images over 498\,h while being recorded with EEG \citep{gifford2022large,grootswagers2022human,xu2024alljoined}, MEG \citep{hebart2023things}, 3T fMRI \citep{hebart2023things,shen2019deep,chang2019bold5000}, or 7T fMRI \citep{allen2022massive}. We systematically evaluate single-trial (real-time-like) decoding performance using the Pearson correlation between the true features of a state-of-the-art image embedding \citep{oquab2023dinov2} and the predictions obtained with brain decoders.

\section{Methods}
\subsection{Problem statement}
\paragraph{Goal} We aim to systematically compare brain-to-image decoding approaches, and identify potential scaling laws, \ie how decoding improves with the type and amount of data. For this, we curate the largest public datasets into a unified benchmark and compare single-trial decoding performance across different experimental setups.

\paragraph{Formalization}
Decoding images from brain signals at the pixel-level is challenging because the brain does not represent images in this feature space \citep{miyawaki2008visual}. Over the years, it has thus become standard to learn to predict latent \emph{embeddings} of images and to use these predictions to condition an image generation model \citep{lin2019dcnn,vanrullen2019reconstructing,ozcelik2023brain,chen2023seeing}. Formally, this approach involves three components:
\begin{itemize}
    \item {\bf Image module} \(\textbf{f}_\theta: \mathbb{R}^{H \times W \times 3} \rightarrow \mathbb{R}^{F}\), to transform image \(I_i\) into a latent embedding \(\mathbf{z}_{i}\),
    \item {\bf Brain module} \(\textbf{g}_\theta: \mathbb{R}^{S \times T} \rightarrow \mathbb{R}^{F}\), to predict an estimate \(\mathbf{\hat{z}}_{i,k}\) of latent \(\mathbf{z}_{i,k}\) from the recording of brain activity \(\mathbf{X}_{i,k}\) in response to the $k^{th}$ presentation of \(I_i\),
    \item {\bf Generating module} \(\textbf{h}_\theta: \mathbb{R}^{F} \rightarrow \mathbb{R}^{H \times W \times 3}\) to transform  \(\mathbf{\hat{z}}_{i,k}\) into \(\hat{I}_{i,k}\) ,
\end{itemize}
where:
\begin{itemize}
    \item \(S\) is the number of channels (M/EEG) or voxels (fMRI),
    \item \(T\) is the number of time points in the brain signal window, or \emph{repetition times} (\emph{TRs}) for fMRI,
    \item \(i \in [1, N]\) indexes the unique images \(I\) with a common size $H\times W \times 3$,
    \item \(k \in [1, K]\) indexes the image presentations \(I_{i,k}\) of the same image \(I_i\),
    \item \(\mathbf{X}_{i,k} \in \mathbb{R}^{S \times T}\) is the brain data within a time window relative to \(I_{i,k}\),
    \item \(\mathbf{z}_i \in \mathbb{R}^{F}\) is the representation of dimension \(F\) of image \(I_i\) obtained with the pretrained image module $\textbf{f}_\theta$.
\end{itemize}

\subsection{Denoising} 
\label{subsec:denoising}

Brain signals are often denoised before they are fed into the brain module $\textbf{g}_\theta$.
For example, it is common to average the \(K\) brain responses to the same image ($\bar{\mathbf{X}}_i = \frac{1}{K} \sum_{k=1}^{K} \mathbf{X}_{i,k}$) or to average the \(K\) predicted embeddings ($\bar{\mathbf{z}}_i = \frac{1}{K} \sum_{k=1}^{K} \hat{\mathbf{z}}_{i,k}$).
For fMRI, it is also common to first fit a Generalized Linear Model (GLM) \citep{friston1995characterizing} on the whole dataset (or on each of its recordings). Generally, GLMs estimate the fMRI response from the convolution of each image-presentation boxcar function with a parameterizable hemodynamic response function (HRF).
However, as the resulting parameters $\hat{\boldsymbol{\beta}}$ cannot be applied in real-time and are typically computed irrespective of train/test splits, we will here focus on predicting images from $\mathbf{X}$ directly. 
More generally, denoising strategies are often paradigm-dependent (\eg the results will depend on the number of repetitions within and across subjects), which can hinder the comparison of decoding performances and limit their transferability to real-time applications. To address this, we primarily focus on single-trial performance without denoising.

\subsection{Brain modules}
\label{subsec:brain_module}
We implement two state-of-the-art architectures to predict image embeddings from brain activity.
A detailed description of the hyperparameter search procedure and architecture configurations is provided in \Cref{app:architectures}. 
For clarity, we compare these architectures to a simple ridge-regularized linear regression trained to predict image embeddings from either M/EEG or fMRI.

\paragraph{Linear model baseline}
The linear model is a ridge regression~\citep{hoerl1970ridge}.  
In practice, we use scikit-learn's \texttt{RidgeCV} \citep{pedregosa2011scikit}, with $\alpha$ selected log-linearly between $10^{-4}$ and $10^8$ and otherwise default parameters.
We train and evaluate on each subject separately.

As compared to linear models, deep learning architectures make it easier to learn on data from multiple individuals at once (\eg with subject-specific layers or embeddings \citep{defossez2022decoding,chehab2022deep}), leveraging cross-subject brain activity patterns into representations that maximally align with the pretrained image representation.

\paragraph{M/EEG deep learning module}
We use the convolutional architecture of \citet{defossez2022decoding,benchetrit2024brain}. 
This architecture includes a spatial attention layer, a subject-specific linear layer, a series of residual dilated convolutional blocks, a temporal aggregation layer, and two projection heads (one for each loss term described below in \Cref{eq:combined_loss}).
In the largest configuration obtained with hyperparameter search, this yields a total of 20.8M parameters.

\paragraph{fMRI deep learning module}
We adapt the convolutional architecture of \citet{scotti2023reconstructing} for (1) multi-subject training and (2) handling BOLD data with a time dimension \footnote{The original architecture is designed to receive GLM $\hat{\boldsymbol{\beta}}$ as input, and thus does not expect a temporal dimension on its input.}.
For this, we first apply a subject-specific linear projection in the spatial dimension, akin to what is done in the M/EEG module and similar to recent work on architectures designed to work with fMRI $\hat{\boldsymbol{\beta}}$ \citep{scotti2024mindeye2}.
Second, a timestep-wise linear spatial projection (\emph{TR layer}) is used to facilitate the extraction of time-varying information.
As in the original architecture, this projection is followed by layer normalization, a GELU non-linearity, dropout (p=0.5), and residual convolutional blocks.
A temporal aggregation layer then pools the temporal dimension, followed by a linear projection. 
Finally, as in the M/EEG module, projection heads map predictions to the different loss terms\footnote{Of note, we use layer normalization and GELU activation in the CLIP head only, as in the original architecture.}.
In its largest configuration, this yields a total of  146.3M parameters.

\paragraph{Training objective}
To learn to predict image embeddings from brain activity, we use a combined retrieval\footnote{We use only the brain-to-image term of the CLIP loss as in \citet{defossez2022decoding,benchetrit2024brain}.} and reconstruction loss, as in \citet{benchetrit2024brain}:

\begin{equation}
\label{eq:clip_loss}
    \mathcal{L}_{CLIP}(\theta) = -\frac{1}{B} \sum_{i=1}^{B} \log \frac{\exp(s(\hat{\vz_i}, \vz_i)/\tau)}{\sum_{j=1}^{B} \exp(s(\hat{\vz_i}, \vz_j)/\tau)}
\end{equation}

\begin{equation}
    \mathcal{L}_{MSE}(\theta) = \frac{1}{NF} \sum_{i=1}^{N} \lVert \vz_i - \hat{\vz}_i \rVert^2_2
\end{equation}

\begin{equation}
\label{eq:combined_loss}
    \mathcal{L}_{Combined} = \lambda \mathcal{L}_{CLIP} + (1-\lambda) \mathcal{L}_{MSE}
\end{equation}

where $s$ is the cosine similarity and $\tau$ is a temperature parameter that we set to 1 in all experiments. Based on early experiments, we fix $\lambda=0.25$ to balance out the contribution of the two loss terms.

\paragraph{Training details}

We train M/EEG and fMRI modules using the Adam optimizer \citep{kingma2014adam} with default parameters ($\beta_1$=0.9, $\beta_2$=0.999) for up to 50 epochs.
The learning rate and batch size were selected as part of the hyperparameter search procedure and differ per neuroimaging device and data regime (\Cref{app:architectures}).
We use early stopping on a validation set obtained by randomly sampling 20\% of the training data, with a patience of 10 epochs, and evaluate the performance of the selected model on a held-out test set.
Models are trained on a single Volta GPU with 32~GB of memory.
We repeat training using three different random seeds for the weight initialization of the brain module, with two exceptions in \Cref{subsec:data_quantity}: first, when analyzing the impact of the number of subjects, we additionally repeat the sampling of subjects three times; second, when analyzing the impact of trial quantity, we instead use two random seeds for weight initialization and two random seeds for subsampling training trials.

\subsection{Image and reconstruction modules}
\label{subsec:image_modules}

We focus our benchmark on the ability to predict, from brain activity, the latent embeddings of a state-of-the-art computer vision model. 
For this, we use DINOv2-giant \citep{oquab2023dinov2}\footnote{\url{https://huggingface.co/facebook/dinov2-giant}} and take the average output token as target for our embedding prediction task ($F=1536$).
DINOv2 image embeddings have shown great transferability and performance on multiple computer vision downstream tasks and yielded high brain-to-image retrieval performance in previous work \citep{benchetrit2024brain}.
We z-score-normalize the latent embeddings of the images of each dataset by using the training set statistics\footnote{Normalization is only applied to the targets of $\mathcal{L}_{MSE}$, as $\mathcal{L}_{CLIP}$ already contains a normalization step through the use of cosine similarity.}.
Additionally, we compare images reconstructed from brain activity using the methodology of \citet{ozcelik2023brain}, for four representative datasets (see description in \Cref{subsec:image_reconstruction}). 

\subsection{Data}
\label{subsec:data}

We use eight publicly available datasets of brain activity recorded in response to image stimuli:
\texttt{Xu2024} (Alljoined) \citep{xu2024alljoined},
\texttt{Grootswagers2022} (THINGS-EEG1) \citep{grootswagers2022human},
\texttt{Gifford2022} (THINGS-EEG2) \citep{gifford2022large},
\texttt{Hebart2023meg} (THINGS-MEG) \citep{hebart2023things},
\texttt{Shen2019} (DeepRecon) \citep{shen2019deep},
\texttt{Hebart2023fmri} (THINGS-fMRI) \citep{hebart2023things},
\texttt{Chang2019} (BOLD5000) \citep{chang2019bold5000} and
\texttt{Allen2022} (Natural Scenes Dataset, or NSD) \citep{allen2022massive}.
The datasets are summarized in \Cref{tab:datasets}.
A detailed description of each dataset is provided in \Cref{sec:datasets}.
The brain imaging data was collected and publicly shared by the authors of each dataset~\citep{xu2024alljoined,grootswagers2022human,gifford2022large,hebart2023things,shen2019deep,chang2019bold5000,allen2022massive}.

For the THINGS-derived datasets (\texttt{Grootswagers2022}, \texttt{Gifford2022}, \texttt{Hebart2023meg}, \texttt{Hebart2023fmri}), we removed from the training set the images whose category was also in the test set to avoid categorical leakage between the train and test splits as in \citet{benchetrit2024brain}.
On \texttt{Allen2022}, we follow previous image decoding work and use only the four (out of eight) subjects that completed all 40 recording sessions.

We subsample the test set of each dataset by randomly selecting 100 unique test images (except for \texttt{Shen2019}, which has only 50 test images available), and keeping all repetitions for these 100 images.
When studying the impact of averaging over multiple repetitions at test time, we also vary the number of available repetitions in the test set, and evaluate decoding on averaged repetitions (either within- or across-subjects).

\begin{table}[]
\centering
\caption{Image decoding datasets used in this study. See \Cref{sec:datasets} for a detailed description of each dataset.}
\label{tab:datasets}
\resizebox{\textwidth}{!}{%
\begin{tabular}{@{}lcrrrrr@{}}
\toprule
\multicolumn{1}{c}{\textbf{Study}} &
  \multicolumn{1}{c}{\textbf{Device}} &
  \multicolumn{1}{c}{\textbf{\# subjects}} &
  \multicolumn{1}{c}{\textbf{\# sessions}} &
  \multicolumn{1}{c}{\textbf{\# unique images}} &
  \multicolumn{1}{c}{\textbf{\# trials}} &
  \multicolumn{1}{c}{\textbf{Time (h)}} \\ \midrule
\citet{xu2024alljoined}                 & \textcolor{eeg}{EEG}       & 8  & 12  & 960   & 43,070   & 11.6  \\
\citet{grootswagers2022human} & \textcolor{eeg}{EEG}        & 48 & 48  & 22,448 & 1,168,416 & 44.2  \\
\citet{gifford2022large}           & \textcolor{eeg}{EEG}        & 10 & 40  & 16,740 & 83,0640  & 79.9  \\
\citet{hebart2023things}      & \textcolor{meg}{MEG}        & 4  & 48  & 22,448 & 98,592   & 46.4  \\
\citet{shen2019deep}                  & \textcolor{fmri3t}{fMRI (3T)} & 3  & 45  & 1,250  & 23,760   & 57.4  \\
\citet{hebart2023things}   & \textcolor{fmri3t}{fMRI (3T)} & 3  & 36  & 8,740  & 29,520   & 42.6  \\
\citet{chang2019bold5000}            & \textcolor{fmri3t}{fMRI (3T)} & 4  & 54  & 4,916  & 18,870   & 55.0  \\
\citet{allen2022massive}         & \textcolor{fmri7t}{fMRI (7T)} & 4  & 160 & 37,000 & 120,000  & 160.4 \\ \midrule
Total         & & 84 & & & 2,332,868 & 497.5 \\ \bottomrule
\end{tabular}
}
\end{table}

\subsection{Preprocessing}
\label{subsec:preprocessing}

\paragraph{M/EEG} We apply minimal preprocessing to M/EEG data following previous work \citep{defossez2022decoding,benchetrit2024brain}.
First, raw data is highpass-filtered above 0.1~Hz and downsampled to 120~Hz.
Each channel is independently normalized using a robust scaler and values outside [-20, 20] are clipped to minimize the impact of large outliers.
Data is then epoched relative to stimulus onset and baseline-corrected by subtracting the channel-wise average value from the pre-stimulus interval.
Epochs always extend to 1~s after stimulus onset, but have different start times $t_0$ based on previous research: -0.1 for \texttt{Grootswagers2022} (as in \citet{grootswagers2022human}), -0.2 for \texttt{Gifford2022} (as in \citet{gifford2022large}) and -0.5 for \texttt{Hebart2024meg} (as in \citet{benchetrit2024brain}).
For \texttt{Xu2024}, we start epochs -0.3~s relative to stimulus onset to use as much of the previous interstimulus interval segments as possible, however, we use the same interval as in \citet{xu2024alljoined} for baseline correction, \ie (-0.05, 0.0).

\paragraph{fMRI}
We use \texttt{fMRIPrep 23.2.0}  \citep{esteban2019fmriprep} with default parameters to process the fMRI datasets into the standard space \texttt{MNI152NLin2009aSym} \citep{fonov_unbiased_2009}.
Each brain volume of the time series is then projected onto the \texttt{fsaverage5} surface \citep{Fischl1999}.  
This yields, for each recording run, a time series of brain volumes of shape $(20484, T)$ where $T$ is the total number of TRs recorded for this run. Following this, we remove low-frequency noise in the fMRI signal using an additional detrending step: we fit a cosine-drift linear model to each voxel in the time series, and subtract it from the raw signal. Each time series is then z-score-normalized.
Finally, data is epoched into windows of five TRs with the following (start, end) times relative to stimulus onset (in seconds): \texttt{Shen2019} (3.0, 13.0), \texttt{Hebart2024fmri} (3.0, 10.5), \texttt{Chang2019} (3.0, 13.0) and \texttt{Allen2022} (3.0, 11.0).

\subsection{Evaluation} 
We evaluate the ability of brain modules to predict $\mathbf{z}_{i,k}$ given $\mathbf{X}_{i,k}$ across different datasets, subjects and numbers of unique image presentations.
To evaluate prediction performance, we compute the average feature-wise Pearson correlation $R = \frac{1}{F} \sum_{f=1}^{F} \text{corr}(\mathbf{z}^{\left(f\right)}, \hat{\mathbf{z}}^{\left(f\right)})$.
Whenever applicable, we also report the standard error of the mean computed across subjects.
Of note, we use the output of the MSE head to evaluate performance as the reconstruction objective $\mathcal{L}_{MSE}$ is conceptually more aligned with the feature-wise Pearson correlation evaluation metric.

\subsection{Scaling laws}
\label{subsec:settings}
We evaluate the scaling behavior of image decoding models by varying data quantity along two axes: (1) number of training trials and (2) number of subjects.

\paragraph{Image quantity analysis}
We focus on single-subject models and vary, in the training set, the number of unique images as well as the number of image repetitions, whenever available (\ie in \texttt{Allen2022}, \texttt{Gifford2022}, \texttt{Xu2024}). 
We repeat this analysis for the first 10 subjects of every dataset.

\paragraph{Subject quantity analysis} 
We vary the number of subjects seen by the models from one to all subjects available in a given dataset.
For this, we compare two additional configurations: first, we use all available data per dataset (\emph{all-trials}) and second, we approximately match the number of trials across datasets (\emph{matched-trials}, described in \Cref{sec:matched_trials}).

\subsubsection{Recording time estimation}
While the number of image presentations is a straightforward measure of data quantity, it does not reflect the longer stimulus presentation times and interstimulus intervals used in some datasets.
For instance, fMRI datasets relied on much longer stimulus onset asynchrony (SOA) durations (\ie the time elapsed between the start of one image presentation and the start of the following image presentation) than M/EEG to account for the slow hemodynamic response.
For instance, the SOA is 10\,s for \texttt{Chang2019} but only 100\,ms for \texttt{Grootswagers2022}.
Therefore, we additionally study scaling laws from the angle of recording duration, as computed by multiplying the number of training trials by the SOA. 

\subsubsection{Cost estimation} 
When building a new dataset, the choice of neuroimaging device and the targeted quantity of data is strongly influenced by the cost of data acquisition.
As cost varies greatly between different neuroimaging devices, it is therefore useful to also study how it relates to decoding performance.
To provide an \emph{approximate} scaling cost for each device, we surveyed publicly available information about neuroimaging data collection services (\Cref{app:cost}). 
Based on this information, hourly cost (in USD) is estimated at \$263 for EEG, \$550 for MEG, \$935 for 3T fMRI and \$1093 for 7T fMRI.
The reader should bear in mind that these are rough estimates and that data acquisition costs can vary significantly between countries and institutions.

\subsection{Image reconstruction}
\label{subsec:image_reconstruction_methods}
The analyses described above focus on the decoding of an image embedding given single-trial or test-time averaged brain signals.
However, it is becoming increasingly common to evaluate decoding pipelines on their ability to reconstruct the original image rather than its latent representation only. Following this approach, we implement an additional generation step for all decoders.

\paragraph{Reconstruction pipeline} Adapted from \citet{ozcelik2023brain}, our reconstruction pipeline consists in using the predicted embedding as input to a pretrained image generation model. We thus re-trained each decoder to predict the three required embeddings, namely CLIP-Image (257 tokens × 768), CLIP-Text (77 tokens × 768), and AutoKL (4 channels × 64 × 64), using the same objective as before (\Cref{eq:combined_loss}). Images are center-cropped and rescaled to 512 × 512 pixels. Following \citet{ozcelik2023brain}, we use a renormalization step before running image generation: we z-score normalize predicted embeddings, then ``de-normalize'' them using the inverse z-score transform, fitted on the training set.
We run diffusion with 50 DDIM steps, a guidance scale of 7.5, a strength of 0.75 and a mixing of 0.4. 
We additionally blur all human faces generated by the model.

For THINGS-derived datasets, the CLIP-Text embedding is obtained from the THINGS-Image database object-category of the stimulus image. For \texttt{Allen2022}, we follow \citet{ozcelik2023brain} and average the CLIP-Text embeddings of the (at most 5) captions of the corresponding image. 

\paragraph{Training details} In decoding experiments, we predict the same embedding $\vz_i$ for the MSE and CLIP heads (\ie the token-average of DINOv2-giant). By contrast, we obtain better reconstruction performance by predicting distinct embeddings on each head. Specifically, for each of the three embedding prompts needed for reconstructing an image, we train a model to predict the full embedding on the MSE head and a pooled version on the CLIP head (for the CLIP-Image model, we pool by using the class-token; for CLIP-Text, the token-average; for the AutoKL model, the channel-average).

This particular choice of pooling for each embedding was found to perform significantly better across studies and devices.
Finally, for simplicity, we showcase the image reconstructions of one representative study per recording device. We focused on studies using the images from THINGS-dataset whenever possible.

\section{Results}

\subsection{Encoding analysis}

To validate the datasets, we first test a standard linear encoding pipeline~\citep{naselaris2011encoding}.
Here, encoding refers to the prediction of brain responses given the image features. For this, we build a time-lag concatenation of image features to predict, with a ridge regression, the amplitude of the neural time series at each time point relative to stimulus onset (see \Cref{sec:encoding}). We use DINOv2 \citep{oquab2023dinov2} to extract image features, as this unsupervised model has been shown to capture representations similar to those of the brain \citep{benchetrit2024brain} (see \Cref{subsec:image_reconstruction} for analyses using other pre-trained image representations).
These linear encoding models are trained for each subject separately. 
Finally, we use the Pearson correlation (R) to evaluate the similarity between the true and predicted brain responses held out from the same subject.

The encoding results confirm that EEG, MEG, 3T and 7T fMRI can be reliably predicted from the features of the images that subjects watched (see \Cref{fig:pipeline}B and \ref{fig:encoding_analysis}). As expected, brain responses to visual stimuli are best predicted in the occipital lobe, host of the visual cortices, but these responses are visible in a distributed set of brain regions. 
Overall, these results confirm that the brain responses to visual stimuli can be modeled, for each of these datasets, from the pretrained embedding of a computer vision model.

\subsection{Linear decoding}
\label{subsec:linear_decoding}

Encoding analyses make it difficult to compare different types of brain recordings, because the space in which they are evaluated (\eg voxels or sensors) varies arbitrarily between datasets. 
Thus, we now turn to linear decoding.
We train and evaluate linear ridge regression decoders at each time step relative to image onset, and evaluate how well image features can be predicted from brain activity patterns across time.

\Cref{fig:stepwise_decoding}A and C show that image embeddings can be maximally decoded 110\,ms and 380\,ms after image onset for EEG and MEG, respectively. 
For fMRI, which captures the slow blood-oxygen-level-dependent (BOLD) response, decoding performance peaks around 4.5-5.2\,s after image onset.
The decoding performance then decreases to chance-level around 750\,ms (EEG), 1,500\,ms (MEG), 7.5\,s (3T fMRI) and 16\,s (7T fMRI) after image onset.

Switching to the \emph{matched-trials} setting to allow comparing datasets using similar numbers of unique images (\Cref{fig:stepwise_decoding_matched_trials}), we observe that linear decoding is best achieved with 7T fMRI (R=0.238), followed by 3T fMRI (R=0.075-0.126), MEG (R=0.057) and EEG (R=0.018-0.033).

\subsection{Decoding with deep learning}
\label{subsec:decoding_deep_learning}

To what extent can these linear decoding scores be improved with deep learning models?
To address this question, we implemented and trained the deep learning pipelines described in \Cref{subsec:brain_module} on sliding windows.
To account for different temporal resolutions across devices, we used 100-ms windows for M/EEG, and a single TR per window for fMRI (1 TR corresponds to 1.5 to 2\,s in the curated datasets).

\begin{figure}
    \centering
    \includegraphics[width=1.0\linewidth]{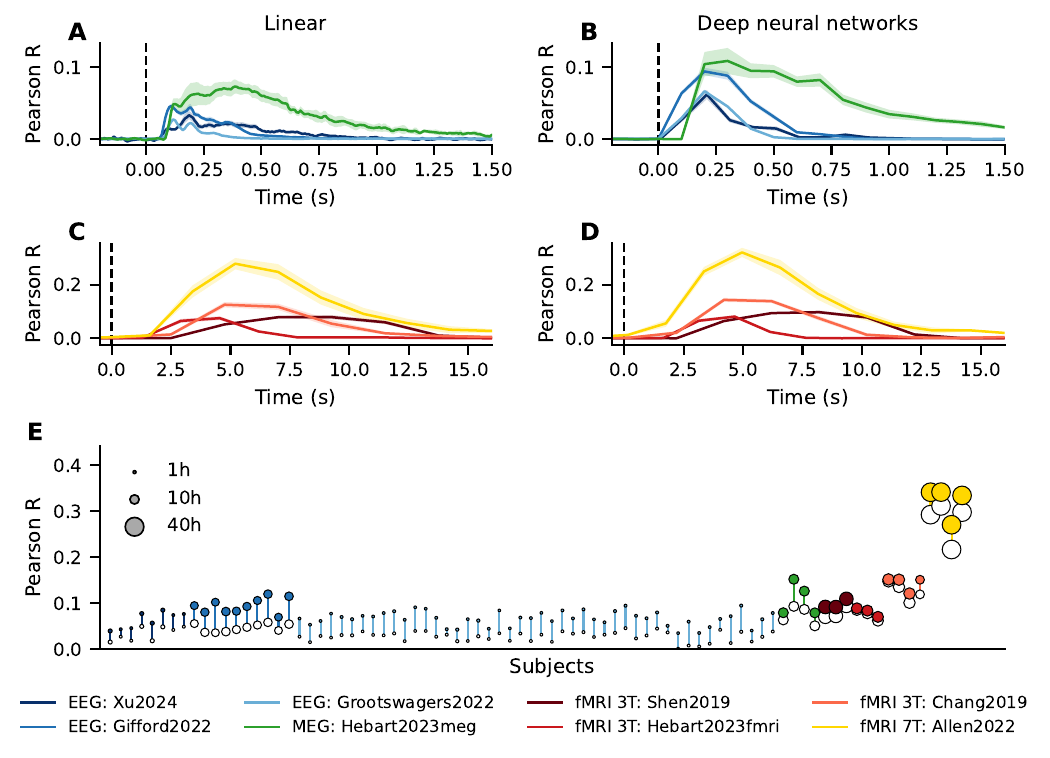}
    \caption{Image decoding analyses show the expected temporal response for all datasets.
    (\textbf{Left}) Subject-specific stepwise decoding using linear ridge regression as a function of time elapsed since image onset ($t=0$).
    (\textbf{Right}) Sliding window decoding using deep learning models trained across subjects on 100-ms windows for M/EEG or 1 TR for fMRI.
    We report the average performance across subjects and show the standard error of the mean with shaded areas or error bars.
    (\textbf{Bottom}) Peak Pearson correlation obtained in the linear (empty circles) and deep learning (filled circles) analyses, for each subject of each study. Circle size indicates the total recording time available for each subject.
    See \Cref{fig:stepwise_decoding_matched_trials} for results in the \emph{matched-trials} setting.
    }
\label{fig:stepwise_decoding}
\end{figure}

These deep learning models reveal a similar decoding dynamic (\Cref{fig:stepwise_decoding}B-D). Critically, we observe a significant improvement over linear baselines (two-sided Wilcoxon signed rank test \citep{Wilcoxon1945} across subjects and datasets, $p < 10^{-14}$; see \Cref{fig:stepwise_decoding}E).
Device performance was ordered similarly to linear models, with EEG and 7T fMRI leading to the worst and best performances, respectively.  
Interestingly however, the \emph{gain} in performance observed between linear and deep models varies across devices: 
1.9-2.4x for EEG, 
1.5x for MEG, 
1.1-1.2x for 3T fMRI, 
and 1.2x for 7T fMRI. 
In other words, the devices that benefit the most from deep learning pipelines appear to be those that are typically associated with lower signal-to-noise ratios.

\subsection{Impact of test-time averaging}
\label{subsec:test_trial_averaging}

In all datasets, test set images were shown multiple times to each subject. This allows improving signal-to-noise ratio and decoding performance, by \eg learning to predict from an averaged response, or averaging single-trial predictions before evaluating metrics.

\begin{figure}
    \centering
    \includegraphics[width=1.0\linewidth]{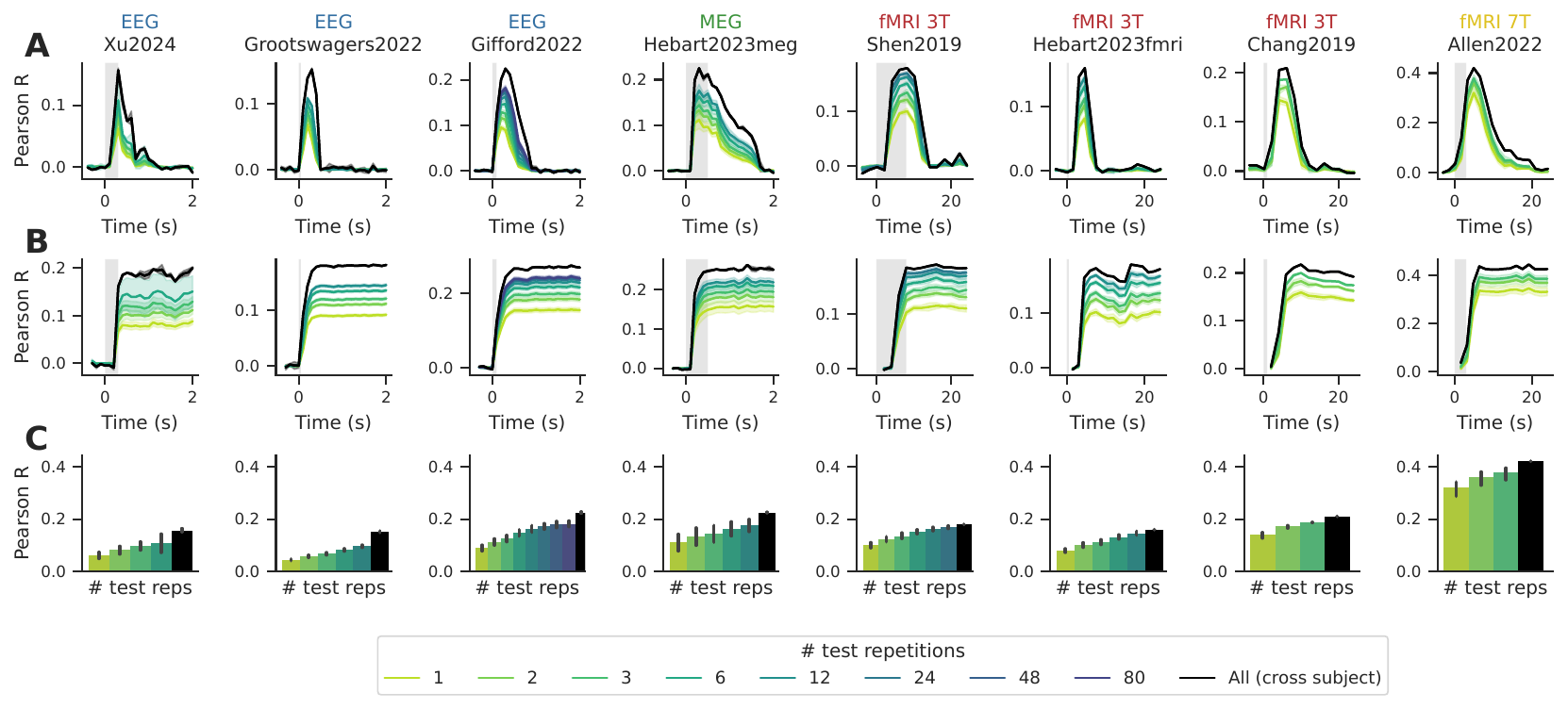}
    \caption{Decoding of the image embedding as a function of time (x-axis) and number of test-time repetitions (color) using deep learning models.
    (\textbf{A}) \textbf{Sliding} window decoding (100~ms for M/EEG; 1~TR for fMRI) and (\textbf{B}) \textbf{growing} window decoding ($t_0$=-0.5 for M/EEG; 0.0 for fMRI) show the expected time-locked response and highlights the consistent improvement obtained by adding test-time image repetitions.
    Grey areas indicate the interval during which images were shown.
    (\textbf{C}) Peak sliding window performance for each dataset.
    Black lines and bars indicate performance obtained when averaging predictions over all repetitions of all subjects for each unique test image.
    We report the average performance across subjects and show the standard error of the mean with shaded areas or error bars.
    We use ``large'' architecture configurations everywhere (\Cref{app:architectures}) except for \texttt{Grootswagers2022} for which the "medium" configuration yielded more stable training dynamics.
    See \Cref{fig:test_trial_averaging_matched_trials} for results in the \emph{matched-trials} setting.
    }
    \label{fig:test_trial_averaging}
\end{figure}

To evaluate whether this multiple-repetition approach effectively scales decoding performance, we systematically vary the number of test image repetitions used for averaging over the test set. 
Results for both sliding- and growing-window models are shown in \Cref{fig:test_trial_averaging}A-B.
For all devices, adding more test repetitions improves decoding performance (Spearman rank correlation of 0.33-0.72, 0.51, 0.58-0.97 and 0.82 for EEG, MEG, 3T and 7T fMRI, respectively).
However, this gain follows a moderate log-linear relationship: the gain in decoding performance rapidly decreases with the increasing number of repetitions.
Further averaging test predictions across subjects (black lines and bars in \Cref{fig:test_trial_averaging}) leads to additional small improvements.
Overall, we observe diminishing returns for all datasets, suggesting test-time averaging may not be ideal to scale decoding performance.

\subsection{Image retrieval}
\label{subsec:image_retrieval}

Next, we evaluate the performance of our decoding pipeline in a retrieval setting: given the predicted DINOv2-giant embedding \(\mathbf{\hat{z}}\) for an image in the test set, we identify the image, among the 100 (50 for \texttt{Shen2019}) unique images in the test set
whose true image embedding is most similar to \(\mathbf{\hat{z}}\).
Of note, the prediction we use is the output of the CLIP head of our model, as it is specifically trained using a retrieval objective ($\mathcal{L}_{CLIP}$).

\begin{figure}
    \centering
    \includegraphics[width=\linewidth]{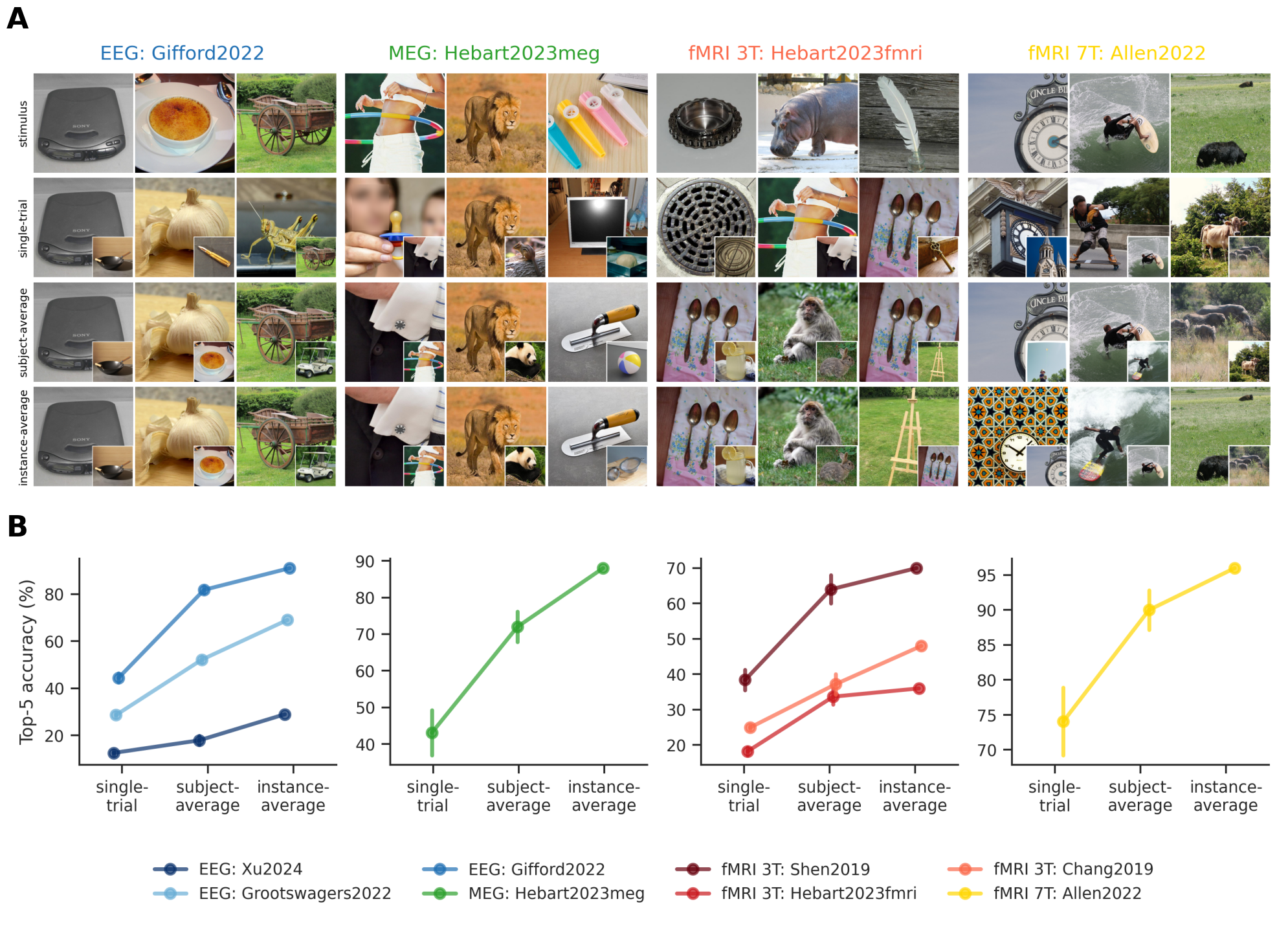}
    \caption{Image retrieval across devices. (\textbf{A})
    For each representative dataset of each device, a sample of three stimulus images showing some of the most convincing retrievals obtained with our approach.
    Ground truth images are shown on the top row.
    Top-1 retrieved images are shown underneath (top-2 image overlaid on bottom right): single-trial brain responses (second row), subject-averaged predictions (third row) and predictions averaged across all subjects (bottom row). (\textbf{B}) Top-5 retrieval accuracies for each dataset and each test-time averaging strategy, grouped by recording device.
    See \Cref{fig:best_retrievals_matched} for the \emph{matched-trials} setting.
    }
\label{fig:best_retrievals}
\end{figure}

\Cref{fig:best_retrievals}A shows a sample of the best retrievals obtained across one representative dataset per device, to align with reconstruction analyses (\Cref{fig:best_generations}).
For all devices, top-1 or top-2 predictions are often correct, and wrong predictions usually share categorical semantic information with the ground truth (\eg animals, inanimate objects, etc.).
Of note, the performance on \texttt{Shen2019} can be partially attributed to the smaller size of its retrieval set (50 vs. 100).

Overall, these results confirm that the deep learning models can accurately identify an image given a pool of candidate images. 
See \Cref{sec:reprod_matched_trials} for comparable results in the \emph{matched-trials} setting.

\subsection{Image reconstruction}
\label{subsec:image_reconstruction}

Image retrieval requires having access to the true image in the test set. To alleviate this constraint, we also evaluate image reconstructions from our decoders. For this, we conditioned the generation of images from the decoded image embeddings, as described in \Cref{subsec:image_reconstruction_methods}. 

\begin{figure}
    \centering
    \includegraphics[width=\linewidth]{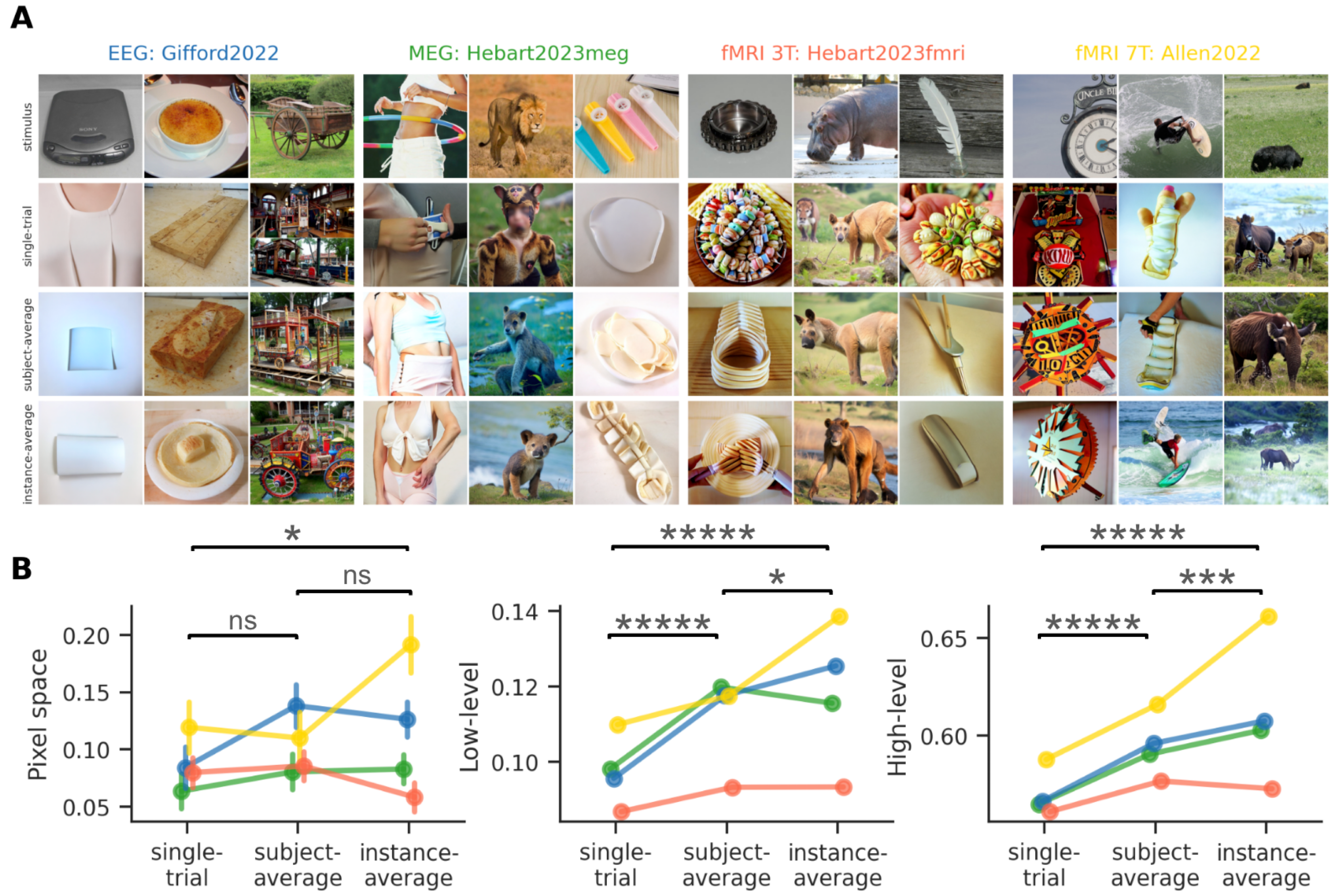}
    \caption{Image reconstruction across devices. (\textbf{A}) For each study, a sample of 3 stimuli showing some of the most convincing reconstructions obtained with our approach: from single-trial brain signals, and two increasingly large aggregations (averaging embedding predictions at subject-level and at instance-level). (\textbf{B}) The comparison of the three image generation metrics PixCorr (pixel space), AlexNet-2-R (low-level, latent space) and CLIP-Final-R (high-level, latent space) for single-trial decoding and aggregations. As expected, performance increases overall when averaging more than one trial, even at subject-level.}
    \label{fig:best_generations}
\end{figure}

\Cref{fig:best_generations} shows a sample of the reconstructions obtained on the same images as in \Cref{fig:best_retrievals}. Overall, the images are never perfectly reconstructed, but nevertheless often share low-level as well as semantic features with the images seen by the subjects. 

To quantify these qualitative observations, we evaluate reconstructions with pixel-, low- and semantic-level metrics, and compare them across decoding approaches (\Cref{subsec:test_trial_averaging}).
Specifically, to evaluate the consistency between a stimulus image $I$ and its reconstruction $\hat{I}$, we select a representation method $\tau$ and compute the Pearson correlation between $\tau(I)$ and $\tau(\hat{I})$. In our study, we choose three different representations for $\tau$, covering both perceptual and semantics aspects of images: (1) the unmodified image itself as an array of pixels, (2) the AlexNet-2 embedding, and (3) the CLIP-final embedding of an image. Averaging these correlations across unique images for each representation $\tau$, we obtain three different metrics that we respectively denote as PixCorr, AlexNet-2-R and CLIP-Final-R. Of note, these last two metrics slightly  differ from the 2-way comparison metrics AlexNet-2 and CLIP-Final, defined in \citet{ozcelik2023brain}. The use of the simpler average correlation (a point-wise metric), instead of the 2-way comparison score, allows us to compare different denoising approaches using a two-sided Wilcoxon signed-rank test, as shown in \Cref{fig:best_generations}B.

The results confirm that the best reconstructions are achieved from 7T fMRI. In addition, the quality of reconstructions increases when averaging embeddings across image repetitions, both qualitatively (\Cref{fig:best_generations}A) and quantitatively (\Cref{fig:best_generations}B).
This is the case when averaging all predicted embeddings for a given image across a single subject (``subject-average''), and even more so when averaging all predicted embeddings for a given image across all subjects (``instance-average'').

\subsection{Comparing decoding performance across data quantities}
\label{subsec:data_quantity}

How does image decoding scale with the amount of brain signals?
To address this question, we train our decoders on increasingly larger subsets of neuroimaging data and measure the corresponding single-trial decoding performance (\ie without test-time averaging).

\begin{figure}
    \centering
    \includegraphics[width=1.0\linewidth]{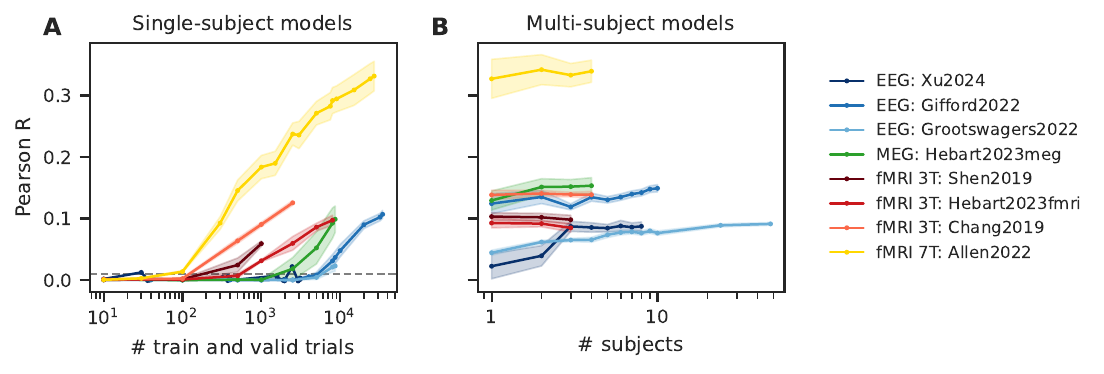}
    \caption{Decoding performance as a function of (\textbf{A}) number of training and validation trials for single-subject models and (\textbf{B}) number of subjects for multi-subject models.
    The horizontal dashed line in (\textbf{A}) indicates the threshold of $R=$0.01 which we use as a comparison point for the number of trials required to perform better than chance.
    Shaded areas represent the standard error of the mean across subjects.
    }
    \label{fig:scaling_trials_and_subjects}
\end{figure}

\paragraph{Scaling trials} First, we consider the scaling of decoding performance \emph{within subjects}.
\Cref{fig:scaling_trials_and_subjects}A shows that different devices require different numbers of trials to reach the same decoding performance. 
A log-linear fit shows that, to reach $R=0.01$, 7T fMRI only requires 57 trials, whereas 3T fMRI requires between 123 and 522 trials, MEG requires 2,3K trials, and EEG requires between 4,9K and 5K trials.

\paragraph{Scaling subjects} Second, we consider the scaling of decoding performance \emph{across subjects} (\Cref{fig:scaling_trials_and_subjects}B).
Results suggest that scaling the number of subjects yields limited improvement in most cases and may even lead to a decrease in performance (\texttt{Shen2019}, \texttt{Hebart2023fmri}).
Although our models are trained and tested on the same subjects, inter-subject variability (see also \Cref{fig:stepwise_decoding}E) appears to harm overall model performance, especially in low-subject regimes, \eg fewer than 10 subjects.
Of note, the \texttt{Grootswagers2022} dataset, with 48 subjects, has the largest number of subjects in our benchmark. Yet, the improvement of 0.004 obtained by doubling the number of subjects from 24 to 48 is not significant (two-sided t-test, $p=0.49$). Scaling the number of EEG subjects does not seem to provide clear benefits for decoding performance.

\begin{figure}
    \centering
    \includegraphics[width=1.0\linewidth]{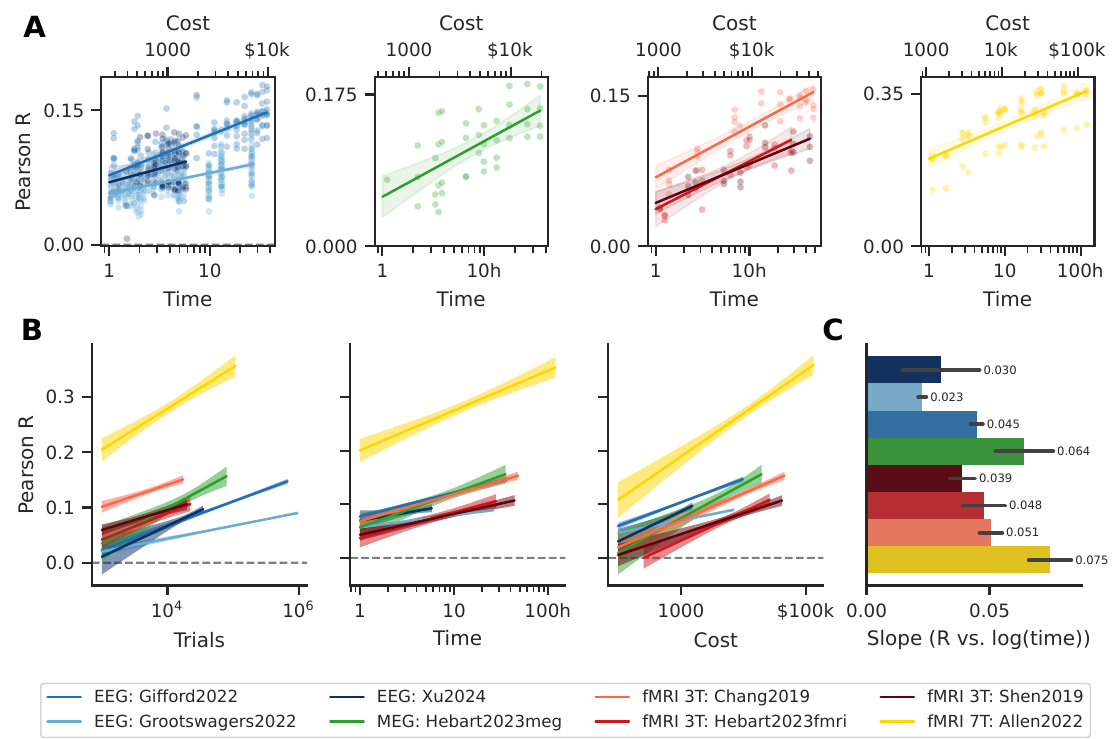}
    \caption{Decoding performance for each dataset as a function of data quantity.
    (\textbf{A}) Per-device performance as a function of image presentation time (bottom x-axis) and of the estimated average data collection cost (top x-axis).
    (\textbf{B}) Performance is shown on the same x-axis (number of trials, recording time in hours or estimated cost in USD) for all devices to facilitate comparison.
    We vary the data quantity by training medium-size models on (i) single subjects, (ii) all subjects but with \emph{matched trials} (see \Cref{sec:matched_trials}) or (iii) all subjects and \emph{all trials}.
    (\textbf{C}) Slope of log-linear models fitted on the number of recording hours.
    Shaded areas and error bars represent the standard error of the mean of the fitted log-linear model parameters.
    }
    \label{fig:data_quantity_analysis}
\end{figure}

\paragraph{Scaling both trials and subjects} Next, we compare data quantity across a mixture of data regimes (\ie number of subjects \emph{and} trials) by considering (1) the total number of training trials across subjects) and (2) the recording duration (\ie the number of hours corresponding to a given number of trials).
As shown in \Cref{fig:data_quantity_analysis}, decoding performance increases with the amount of data across all datasets.
This increase follows a log-linear trend, whose slope and intercept depends on the recording device. 
For example, focusing on models fitted on the number of recording hours (\Cref{fig:data_quantity_analysis}B, middle column), EEG has a slope of 0.045 ($\pm$0.003 standard error of the mean), MEG of 0.064 ($\pm$0.011), 3T fMRI of 0.048 ($\pm$0.009) and 7T fMRI of 0.075 ($\pm$0.009).

Strikingly, 7T fMRI (\texttt{Allen2022}) yields the best decoding performance across the board, both in terms of number of trials and hours of recording.
7T fMRI aside, the device that best scales decoding performance depends on what factor (trials, time, cost) is considered (\Cref{fig:data_quantity_analysis}B).
When considering the number of trials, 3T fMRI either outperforms (\texttt{Chang2019}) or works similarly to MEG with a similar number of trials, while EEG lags behind.
When considering the amount of recording time, we find instead that fMRI 3T (\texttt{Chang2019}), MEG and EEG with trial scaling (\texttt{Gifford2022}) yield similar performance for the same number of hours.
When considering the potential of scaling, MEG has the largest slope after 7T fMRI (\Cref{fig:data_quantity_analysis}B), suggesting it may be a promising avenue to scale decoding performances. 
Finally, we do not currently find evidence of a saturation effect: decoding performance does not appear to plateau after a specific quantity of data.

\subsection{Estimating cost of data acquisition}

The possibility of scaling does not solely depend on performance, but also on the cost of data acquisition.
To address this issue, we estimate the cost associated with each type of device, by using publicly available information (see \Cref{app:cost}), and modeled the log-linear relationship between cost and decoding performance.
Fitted models are shown in \Cref{fig:data_quantity_analysis}B (last column).
We can then estimate the cost or the gain associated with different hypothetical objectives.
First, with these cost estimates, we retrospectively infer that the acquisition of the present datasets costs
\$1.5k for \texttt{Xu2024},
\$6.9k for \texttt{Grootswagers2022},
\$9.7k for \texttt{Gifford2022},
\$19.4k for \texttt{Hebart2023meg},
\$41.1k for \texttt{Shen2019},
\$26k for \texttt{Hebart2023fmri},
\$44.8k for \texttt{Chang2019}, and
\$131.2k for \texttt{Allen2022}.
Second, according to log-linear models, a budget of \$131.2k (\ie the estimate for the ultra-high field fMRI dataset) would lead to Pearson correlations of 0.123-0.197 for EEG, 0.210 for MEG, 0.122-0.172 for fMRI 3T and 0.363 for fMRI 7T (the actual maximum Pearson R for fMRI 7T is 0.372).

These estimates highlight the fact that despite the considerable difference between 7T fMRI and other modalities, the high cost and slow temporal resolution of 7T fMRI may not lead to the most effective path to scaling the decoding of images from brain activity.

\section{Discussion}

We aimed to characterize the factors which are critical to improving brain-to-image decoding performance. For this, we conducted a comprehensive comparison of decoding pipelines on the largest benchmark to date encompassing 84 subjects who watched 2.3M natural images over 498\,h while their brain activity was recorded with EEG, MEG, 3T fMRI or 7T fMRI. To ensure meaningful comparisons, we focused on a unified preprocessing and modeling pipeline, evaluated decoding on single-trial and controlled for the amount and type of training data.

\subsection{Contributions}
This work provides three main contributions.

\paragraph{Decoding performance and deep learning gain}
First, both linear and deep learning models highlight the importance of recording devices: as expected, when the size of the training sets are similar, 7T fMRI leads to the best results, followed by 3T fMRI, MEG and ultimately EEG. 
However, the decoding \emph{gain} enabled by deep learning algorithms, as compared to linear models, unexpectedly appears to benefit the noisiest devices, namely EEG and MEG (\Cref{fig:stepwise_decoding}). 
We speculate that the noise associated with brain recordings may have specific spatial (sensors or voxels) and temporal structures that can only be separated in a latent space. 
If confirmed, this hypothesis would highlight the importance of developing, in the future, foundational models of brain activity, to automatically perform such separation between brain signals and recording noise (\eg \citet{thomas2022self,ortega2023brainlm,yang2024biot,yuan2024brainwave}). 

\paragraph{Scaling laws}
Second, and in spite of analyzing the largest amount of brain responses to images to date, we do not observe any plateau of decoding performance as the amount of training data increases. 
Rather, the log-linear scaling laws presently observed strengthens previous findings \citep{antonello2024scaling,bonnasse2024fmri,defossez2022decoding,benchetrit2024brain}. Together, these results suggest that the decoding of brain activity may be most simply improved by recording more data. Interestingly, not all data regimes are equivalent in that regard. In particular, our results show that the amount of \emph{within-subject} recordings steadily improves decoding, whereas the amount of \emph{across-subject} recordings lead to modest, if any, improvements. This result emphasizes the importance of inter-individual differences when it comes to neural representations and brain activity patterns. While additional research incorporating large databases across many more subjects \citep{vanessen2021human,nastase2021narratives,schoffelen2019subject} remains necessary, this result suggests that future efforts may benefit most from focusing on building datasets with a few subjects recorded over many sessions \citep{allen2022massive,hebart2023things,armeni2022hour}.
Finally, the study of scaling laws often considers the impact of data size in relation to the size of the model. The systematic exploration of increasingly large architectures remains an open question~\citep{kaplan2020scaling,hoffmann2024training}.

\paragraph{Beyond performance: the importance of time and cost}
Third, the present comparisons highlight that decoding performance should not be the sole factor to consider when deciding which device to use for data collection. As illustrated in \Cref{fig:data_quantity_analysis}, the temporal resolution and the costs associated with fMRI is such that, depending on the budget, and/or the targeted decoding performance, the most cost- and time-efficient route may not necessarily be in favor of MRI devices.
For instance, when comparing performance for equivalent numbers of trials, 3T fMRI is better or equivalent to MEG, which itself is better than EEG.
However, if we look instead at recording duration, MEG and EEG (\texttt{Gifford2022} and \texttt{Xu2024}) reach better performance than two of the three 3T fMRI datasets.

\subsection{Limitations}

\paragraph{Experimental protocols}
Our main scaling law analysis (\Cref{fig:data_quantity_analysis}) compares devices by subsampling their corresponding datasets. However, each dataset is also marked by differences in experimental paradigms. 
For instance, while \texttt{Gifford2022}, \texttt{Grootswagers2022}, \texttt{Hebart2023meg} and \texttt{Hebart2023fmri} used pictures from the THINGS database~\citep{hebart2019things}, \texttt{Allen2022} and \texttt{Xu2024} used images from COCO \citep{Lin2014microsoft}. 
Yet, different image datasets may imply different biases (\eg category bias, field of view, scene- vs. object-centered), which in turn could impact decoding performance~\citep{Shirakawa2024}. 
In addition, the duration of the image presentations, and of the pause in-between trials, varies across studies, in part to accommodate the temporal resolution of the corresponding brain recording device (\eg SOA of 100\,ms for \texttt{Grootswagers} vs. 10\,s for \texttt{Chang2019}).
It will be crucial for the research community to collect additional datasets so as to formally evaluate the impact of stimulus design choices on brain decoding performance.

\paragraph{Suboptimal image reconstructions}
The present study focuses on a unified pipeline to decode images from single-trial brain responses. This choice, motivated by real-time-like evaluation, may lead to suboptimal decoding performances.
First, unlike many fMRI studies, we do not systematically explore the variety of denoising strategies. In particular, many fMRI studies rely on ``beta-values'', \ie statistical estimates optimized to specifically isolate the brain response to each image (\Cref{subsec:denoising}). This decision stems from the fact that the results of GLMs (1) are not usable in real-time, (2) vary with the number of repetitions and (3) risk mixing the train/test signals, since they are applied before such splitting. Yet, the resulting brain patterns have higher signal-to-noise ratio, and thus lead to better decoding performances~\citep{ozcelik2023brain}. 
Second, most recent decoding studies have made use of models pretrained on fMRI and/or EEG, with demonstrable improvements~\citep{chen2023seeing}.
Third, we here focus on two state-of-the-art architectures, for M/EEG on the one hand \citep{benchetrit2024brain,defossez2022decoding}, and 3T and 7T fMRI on the other hand \citep{scotti2023reconstructing}. However, other architectures and optimizations have been proposed too \citep{ozcelik2022reconstruction, shen2024neurovision, yang2024biot, yuan2024brainwave}. 
Finally, image reconstruction appears to continue improving from the continuous development of generative image models \citep{scotti2024mindeye2}. 
Overall, the continuously-developed architectural, preprocessing, and modeling tricks employed in the field should be prioritized when optimizing for reconstruction quality.

\subsection{Ethical considerations}
The improvement in brain decoding methodology has raised ethical discussions \citep{poldrack2017neuroscience}. While the present study confirms that we can decode images from brain activity, it is restricted to the decoding of visual \emph{perception}, \ie exogenously-elicited representations that experimenters can easily control and repeat. For \emph{endogenous} representations, such as imagination, recall, internal reasoning, etc., current studies show that decoding can only achieve statistical significance if subjects actively engage in a controlled task, and that the decoding performances are effectively mediocre \citep{tang2023semantic,horikawa2013neural}. Consequently, these results suggest that it will not be possible to decode, from neuroimaging, spontaneous train-of-thoughts in real-time and without the consent of the subjects. While these technical hurdles certainly provide a degree of security against the misuse of brain decoding, they also effectively limit the feasibility of applying these approaches in clinical settings, where brain-lesioned patients may benefit from brain-computer-interfacing technology.

\subsection{Conclusion}
Overall, the present study seeks to contribute to the maturation of neuroscience: as our discipline continues to produce increasingly larger datasets~\citep{markiewicz2021openneuro, gorgolewski2016brain} and extensive research findings~\citep{yarkoni2011large, dockes2020neuroquery}, comprehensive benchmarking is going to be an essential tool for modeling and understanding the neural representations of human cognition.


\clearpage
\newpage
\bibliographystyle{assets/plainnat}
\bibliography{main}

\clearpage
\newpage
\beginappendix

\newcommand{\beginsupplement}{
    \setcounter{table}{0}
    \renewcommand{\thetable}{S\arabic{table}}%
    \setcounter{figure}{0}
    \renewcommand{\thefigure}{S\arabic{figure}}%
    \setcounter{equation}{0}
    \renewcommand{\theequation}{S\arabic{equation}}%
}
\beginsupplement

\section{Detailed description of image decoding datasets}
\label{sec:datasets}

\subsection{\texttt{Xu2024} (Alljoined)}
\textbf{Device.} EEG was recorded at 512~Hz using a 64-channel BioSemi ActiveTwo system.
\textbf{Subjects.} Eight subjects (two females and six males, mean age of 22 $\pm$ 0.64 years old) underwent one or two one-hour sessions, for a total of 12 sessions.
\textbf{Stimuli.} A total of 960 natural images were selected from the shared set of test images used in \citet{allen2022massive} (initially taken from MS-COCO~\citep{Lin2014microsoft}).
Each image was shown up to four times per subject across sessions. 
Images were presented for 300~ms, followed by a 300~ms blank screen. 
We set aside 20\% of the unique 960 images to use as test set.

\subsection{\texttt{Grootswagers2022} (THINGS-EEG1)}

\textbf{Device.} EEG was recorded at 1~kHz using a 64-channel BrainVision ActiChamp system, referenced to Cz.
\textbf{Subjects.} 50 healthy volunteers (36 females and 14 males, mean age of 20.44 $\pm$ 2.72 years old) underwent a single one-hour session each.
We discard the last two subjects as their recordings contained a different number of EEG channels.
\textbf{Stimuli.} A total of 22,448 unique images from the THINGS dataset~\citep{hebart2019things} were shown across sessions for each subject. Out of these, 200 images from distinct categories were selected to form the test set and were shown 12 times to each subject. Each image was presented for 50~ms, followed by a 50-ms fixation screen.
Of note, test set annotations are not available for subjects 1 and 6. As a result, while data from these two subjects is included in our training sets, models are not evaluated on them.  

\subsection{\texttt{Gifford2022} (THINGS-EEG2)}
\textbf{Device.} EEG was recorded at 1~kHz using a 64-channel BrainVision ActiChamp system, referenced to Fz.
\textbf{Subjects.} Ten healthy volunteers (eight females and two males, mean age of 28.5 $\pm$ 4 years old) each underwent four sessions of about 1.6~h each.
\textbf{Stimuli.} A total of 16,740 unique images from the THINGS dataset~\citep{hebart2019things} were shown across sessions for each subject.
Among these images, 200 images from distinct categories were selected to form the test set. Training set images were shown four times to each subject, while test set images were shown 80 times to each subject. Each image was displayed for 100~ms, followed by a fixation period of 100~ms. 

\subsection{\texttt{Hebart2023meg} (THINGS-MEG)}
\textbf{Device.} MEG was recorded with a 275-channel CTF system which incorporates a whole-head array of 275 radial 1st order gradiometer SQUID channels sampled at 1,200~Hz. 
\textbf{Subjects.} Four healthy volunteers (two females and two males, mean age of 23.25 year old) each underwent 12 sessions of about one hour each. 
\textbf{Stimuli.} A total of 22,448 unique images from the THINGS dataset~\citep{hebart2019things} were shown across sessions for each subject.
Among these images, 200 images from distinct categories were selected to form the test set and shown 12 times to each subject. 
Each image was displayed for 500~ms, followed by a fixation period varying from 800 to 1,200~ms. 

\subsection{\texttt{Shen2019} (DeepRecon)}

\textbf{Device.} Functional MRI was obtained using a 3 Tesla Siemens MAGNETOM Verio scanner. The imaging used a T2*-weighted gradient-echo EPI multi-band pulse sequence recording the entire brain at TR=2~s (76 slices, slice-thickness 2~mm, slice gap 0~mm and a field-of-view of 192×192~mm).
\textbf{Subjects.} To allow fair comparison with the other datasets included in our study, which contain only natural images, we restrict our analysis to the training and test natural-image sessions. In this context, three healthy volunteers (one female and one male, age range 23 to 33 years old) each underwent 18 fMRI sessions of up to 2 hours each.
\textbf{Stimuli.} The subjects saw a total of 1,250 unique images sampled from ImageNet \citep{deng2009imagenet}. During training sessions (which form our training set), 1,200 unique images were presented 5 times (a total of 6,000 training trials). During test sessions (which form our test set), 50 unique images were shown 24 times each (a total of 1,200 test trials). Each image was displayed for 8~s, with no rest between consecutive image presentations.

\subsection{\texttt{Hebart2023fmri} (THINGS-fMRI)}
\textbf{Device.} Functional MRI was recorded with a 3 Tesla Siemens Magnetom Prisma scanner and a 32-channel head coil, using a repetition time (TR) of 1.5~s. The resolution consisted of the whole-brain with 2~mm isotropic resolution (60 axial slices, 2 mm slice thickness without slice gap, a matrix size of 96×96 and a field-of-view of 192×192~mm).
\textbf{Subjects.} Three healthy volunteers (two females and one male, mean age of 25.33 years old) each underwent 12 sessions of about one hour each.
\textbf{Stimuli.} A total of 8,740 unique images from the THINGS database 
\citep{hebart2019things} were shown across sessions for each subject.  
Among these images, 100 images from distinct categories were selected to form the test set and shown 12 times to each subject. 
Each image was displayed for 500~ms, followed by a fixation period of 4~s.

\subsection{\texttt{Chang2019} (BOLD5000)}
\textbf{Device.} Functional MRI was obtained using a 3 Tesla Siemens Verio MR scanner with a 32-channel phased array head coil. The imaging used a T2*-weighted-gradient recalled-echo EPI multi-band pulse sequence, sampled at TR=2~s and captured with a resolution of 2×2~mm (69 slices co-planar with the AC/PC line, slice-thickness 2~mm, slice gap 0~mm, matrix size 106×106 and a field-of-view of 212×212~mm).
\textbf{Subjects.}~Four healthy volunteers (three females and one male, age range 25 to 27 years old) each participated in 15 sessions, except one who  completed 9 sessions only. Each session lasted 1.5~hours.
\textbf{Stimuli.} The subjects who went through all 15 sessions saw a total of 4,916 unique images sampled from three datasets: Scene UNderstanding \citep{xiao2010sun}, MS-COCO \citep{Lin2014microsoft} and ImageNet \citep{deng2009imagenet}. A subset of 112 images were shown 4 times (or 2, 3 or 5 times for a small sample of images) to each subject and were selected to form the test set. Each image was presented for 1~s, followed by a fixation cross displayed for 9~s.

\subsection{\texttt{Allen2022} (Natural Scenes Dataset, or NSD)}
\textbf{Device.} Functional MRI was recorded with a 7 Tesla Siemens Magnetom passively shielded scanner and a single-channel-transmit, 32-channel-receive RF head coil (Nova Medical) sampled at TR=1.6~s, using a gradient-echo EPI at 1.8~mm isotropic resolution with whole-brain coverage (84 axial slices, slice thickness 1.8~mm, slice gap 0~mm, a matrix size 120×120 and a field-of-view of 216×216~mm).
\textbf{Subjects.} Eight healthy volunteers (six females and two males, mean age range 19 to 32 years old) each underwent 30-40 fMRI sessions of about one hour each. Following previous work on this dataset, we use only data from the subjects who completed the full 40 recording sessions (namely subjects 1, 2, 5, 7). 
\textbf{Stimuli.} A total of 73,000 natural images from the MS-COCO dataset~\citep{Lin2014microsoft} were shown across subjects and sessions. Each subject saw a total of 10,000 unique images (each repeated 3 times) across 40 sessions. 
Of these, 9,000 images were selected for training, while a common set of 1,000 images seen by all subjects was used for the test set. Each image was shown for 3~s, with a 1~s blank interval between consecutive image presentations.

\section{Matching dataset sizes}
\label{sec:matched_trials}

Image decoding datasets vary in size across several dimensions including the numbers of subjects, the amount of recordings per subject, the number of unique images used as well as the number of repetitions of each unique image. 
To minimize the impact of these factors for the comparison of different datasets and devices, we define a \emph{matched-trials} configuration, where datasets are downsampled to match as closely as possible the size of \texttt{Hebart2023fmri}, the smallest of the THINGS-derived datasets.
We report the numbers of unique images and presentation trials in the \emph{matched-trials} and \emph{all-trials} data configurations for each dataset in \Cref{tab:trial_settings}.

For the neuroimaging datasets based on the THINGS images~\citep{hebart2019things} (\texttt{Grootswagers2024}, \texttt{Gifford2022}, \texttt{Hebart2023meg} and \texttt{Hebart2023fmri}), we use 7,428 unique images (\ie the number of training images in \texttt{Hebart2023fmri} after removing test set categories from the original training set; see \Cref{subsec:data}) for the training set, each presented only once.
In \texttt{Gifford2022}, contrarily to other datasets, each training image was presented multiple times. 
Therefore, we sample a unique presentation of each training image per subject.
Since the ultra-high field fMRI dataset (\texttt{Allen2022}) does not make use of the THINGS images, we randomly select 7,428 unique images \emph{per subject} to build a training set (with a single presentation per image).
\texttt{Xu2024}, \texttt{Shen2019} and \texttt{Chang2019} contain fewer image presentations than \texttt{Hebart2023fmri}, therefore we do not downsample them and keep all available training examples.
Finally, for each dataset, we build a test set by randomly selecting 100 unique images (50 for \texttt{Shen2019}) from the original test splits. This test set is used in both \emph{matched} and \emph{all} trial configurations.

\begin{table}[]
\centering
\caption{Description of the data quantity configurations. The number of trials corresponds to the number of total images presentations available in a configuration when including all subjects and repetitions.}
\label{tab:trial_settings}
\begin{tabular}{@{}lrrrrrr@{}}
\toprule
                          & \multicolumn{2}{c}{\textbf{Matched trials}} & \multicolumn{2}{c}{\textbf{All trials}} & \multicolumn{2}{c}{\textbf{Both configurations}} \\ \midrule
\multicolumn{1}{c}{\textbf{Study}} &
  \multicolumn{1}{c}{\textbf{\begin{tabular}[c]{@{}c@{}}\# train+valid\\ unique images\end{tabular}}} &
  \multicolumn{1}{c}{\textbf{\begin{tabular}[c]{@{}c@{}}\# train+valid \\ trials\end{tabular}}} &
  \multicolumn{1}{c}{\textbf{\begin{tabular}[c]{@{}c@{}}\# train+valid \\ unique images\end{tabular}}} &
  \multicolumn{1}{c}{\textbf{\begin{tabular}[c]{@{}c@{}}\# train+valid \\ trials\end{tabular}}} &
  \multicolumn{1}{c}{\textbf{\begin{tabular}[c]{@{}c@{}}\# test unique \\ images\end{tabular}}} &
  \multicolumn{1}{c}{\textbf{\begin{tabular}[c]{@{}c@{}}\# test \\ trials\end{tabular}}} \\ \midrule
\texttt{Xu2024}           & 777                  & 34,868               & 777                & 34,868              & 100                   & 4,472                    \\
\texttt{Grootswagers2022} & 7,428                & 353,172              & 19,848             & 943,892             & 100                   & 55,200                   \\
\texttt{Gifford2022}      & 7,428                & 300,145              & 16,540             & 668,400             & 100                   & 81,091                   \\
\texttt{Hebart2023meg}    & 7,428                & 29,712               & 19,848             & 79,392              & 100                   & 4,800                    \\
\texttt{Shen2019}         & 1,200                & 19,800               & 1,200              & 19,800              & 50                    & 3,960                    \\
\texttt{Hebart2023fmri}   & 7,428                & 22,284               & 7,428              & 22,284              & 100                   & 3,600                    \\
\texttt{Chang2019}        & 4,803                & 17,255               & 4,803              & 17,255              & 100                   & 1,422                    \\
\texttt{Allen2022}        & 29,712               & 89,136               & 36,000             & 108,000             & 100                   & 1,200                    \\ \bottomrule
\end{tabular}
\end{table}

\section{Encoding analysis}
\label{sec:encoding}

\begin{figure}
    \centering
    \includegraphics[width=0.9\linewidth]{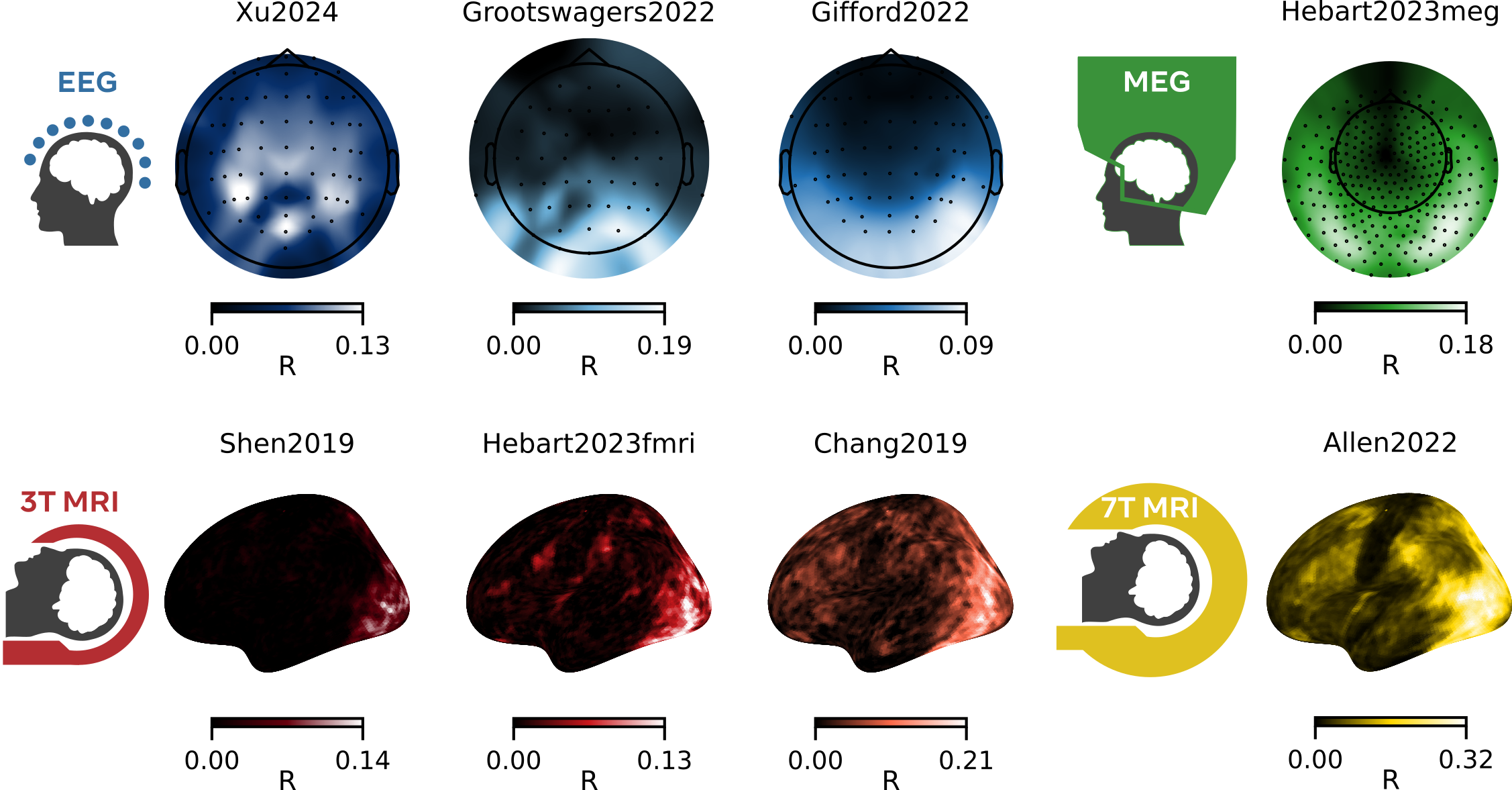}
    \caption{Encoding models trained to predict each M/EEG channel or fMRI voxel from the presented images yield the expected spatial response over the occipital region as measured with Pearson correlation.
    }
    \label{fig:encoding_analysis}
\end{figure}

We perform encoding analyses \citep{king2020encoding} to verify that the image decoding datasets (\Cref{tab:datasets}) capture the expected spatial response over the occipital cortex.
We extract the DINOv2-giant average token of the output layer (see \Cref{subsec:image_modules}) for the images of each dataset.
For each subject of each dataset, we build a collection of image latent and brain response pairs, where we use the brain response at a fixed time $t_{enc}$ after stimulus onset (picked using the approximate maximal response seen in \Cref{fig:stepwise_decoding}A and B).
We use $t_{enc}=0.2$\,s for M/EEG datasets, and dataset-specific offsets for fMRI datasets (\texttt{Shen2019}: 9.5\,s, \texttt{Hebart2023fmri}: 4.5\,s, \texttt{Chang2019}: 5.0\,s and \texttt{Allen2022}: 5.5\,s).
Linear ridge regression encoders (\texttt{RidgeCV} with \texttt{alpha} sampled log-linearly between $10^{-12}$ and $10^{22}$) are then trained to map image latents to the response of a single M/EEG channel or single fMRI voxel.
We use Pearson correlation between ground truth and predicted brain responses on a held-out test set (2-fold cross-validation) to evaluate the quality of the encoding.
Finally, correlation values are averaged across subjects within each dataset, and the average is plotted on topographical maps and inflated brain volumes (\Cref{fig:pipeline}C and \ref{fig:encoding_analysis}).

\section{Hyperparameter search and brain module configurations}
\label{app:architectures}

We ran a random hyperparameter search to identify optimal architecture configurations for the different brain devices under different data regimes.
We picked one representative dataset per brain device family (EEG: \texttt{Gifford2022}, MEG: \texttt{Hebart2023meg}, fMRI: \texttt{Hebart2023fmri}) and define two data regimes: ``large'' (all subjects, all trials) and ``medium'' (one subject, all trials).
To perform hyperparameter search on a given dataset, we further split the existing training set by randomly sampling 20\% of the image categories and using the corresponding examples as an inner test set on which model performance is to be compared.
For both data regimes, we then randomly sampled values for the following hyperparameters:
\begin{itemize}
    \item Batch size: uniform over $\{32, 64, 128, 256, 512\}$
    \item Learning rate: uniform over $[3\times10^{-5}, 3\times10^{-4}, 3\times10^{-3}]$
\end{itemize}
For the M/EEG module only:
\begin{itemize}
    \item Number of convolutional blocks: uniform over $\{0, 1, 2, 3, 4, 5\}$
    \item Hidden size: log-uniform over $[32, 512]$
    \item Backbone output size: log-uniform over $[64, 2048]$
\end{itemize}
For the fMRI module only:
\begin{itemize}
    \item Hidden size: log-uniform over $[32, 2048]$
    \item Number of blocks: uniform over $\{0, 1, 2, 3\}$
    \item CLIP head: with/without
\end{itemize}
Sampling of hyperparameters was repeated a total of 75 times for each brain device and data regime. Moreover, the search was repeated three times (on three different random subjects) when searching on the ``medium'' data regime.
Finally, the configuration yielding the best Pearson R on the inner test set was selected to be used in the different experiments.
The chosen configurations for EEG, MEG and fMRI modules are presented in \Cref{tab:eeg-hps}, \ref{tab:meg-hps} and \ref{tab:fmri-hps}.
The resulting architectures are presented in \Cref{tab:eeg-architecture}, \ref{tab:meg-architecture} and \ref{tab:fmri-architecture}.

\begin{table}[]
\centering
\caption{Results of hyperparameter search on EEG data.}
\label{tab:eeg-hps}
\begin{tabular}{@{}lrr@{}}
\toprule
\multicolumn{1}{c}{\textbf{Hyperparameter}} & \multicolumn{1}{c}{\textbf{Medium}} & \multicolumn{1}{c}{\textbf{Large}} \\ \midrule
\# convolutional blocks & 2        & 4        \\
Hidden size            & 50       & 396      \\
Backbone output size   & 152      & 1411     \\
Batch size             & 32       & 256      \\
Learning rate          & $3 \times 10^{-4}$ & $3 \times 10^{-4}$ \\ \bottomrule
\end{tabular}
\end{table}

\begin{table}[]
\centering
\caption{Results of hyperparameter search on MEG data.}
\label{tab:meg-hps}
\begin{tabular}{@{}lrr@{}}
\toprule
\multicolumn{1}{c}{\textbf{Hyperparameter}} & \multicolumn{1}{c}{\textbf{Medium}} & \multicolumn{1}{c}{\textbf{Large}} \\ \midrule
\# convolutional blocks & 4        & 5       \\
Hidden size            & 181      & 442      \\
Backbone output size   & 564      & 1526     \\
Batch size             & 64       & 512      \\
Learning rate           & $3 \times 10^{-4}$ & $3 \times 10^{-4}$ \\ \bottomrule
\end{tabular}
\end{table}

\begin{table}[]
\centering
\caption{Results of hyperparameter search on fMRI data.}
\label{tab:fmri-hps}
\begin{tabular}{@{}lrr@{}}
\toprule
\multicolumn{1}{c}{\textbf{Hyperparameter}} & \multicolumn{1}{c}{\textbf{Medium}} & \multicolumn{1}{c}{\textbf{Large}} \\ \midrule
Hidden size   & 553      & 1552     \\
\# blocks      & 2        & 0       \\
CLIP head    & No        & Yes     \\
Batch size    & 256      & 64       \\
Learning rate  & $3 \times 10^{-4}$ & $3 \times 10^{-3}$ \\ \bottomrule
\end{tabular}
\end{table}

\begin{table}[]
\centering
\caption{Description of the brain module architectures used with EEG data. We provide the input and output shapes for each layer, as well as the corresponding number of parameters.}
\label{tab:eeg-architecture}
\begin{tabular}{@{}lllrllr@{}}
\toprule
                             & \multicolumn{3}{c}{\textbf{Medium}}  & \multicolumn{3}{c}{\textbf{Large}}     \\ \midrule
\multicolumn{1}{c}{\textbf{Layer}} &
  \multicolumn{1}{c}{\textbf{Input}} &
  \multicolumn{1}{c}{\textbf{Output}} &
  \multicolumn{1}{c}{\textbf{\# params}} &
  \multicolumn{1}{c}{\textbf{Input}} &
  \multicolumn{1}{c}{\textbf{Output}} &
  \multicolumn{1}{c}{\textbf{\# params}} \\ \midrule
Spatial attention            & (64, 144)   & (270, 144)  & 552,960   & (64, 144)    & (270, 144)  & 552,960    \\
Linear projection            & (270, 144) & (181, 144)  & 49,051    & (270, 144)  & (442, 144)  & 119,782    \\
Subject layer                & (181, 144) & (181, 144)  & 32,761    & (442, 144)  & (442, 144)  & 1,953,640  \\
Residual dilated conv blocks & (181, 144) & (181, 144)  & 1,578,320 & (442, 144)  & (442, 144)  & 11,739,520 \\
1x1 conv block               & (181, 144) & (564, 144)  & 270,616   & (442, 144)  & (1526, 144) & 1,742,122  \\
Temporal aggregation         & (564, 144) & (564, 1)    & 145       & (1526, 144) & (1526, 1)   & 145        \\
MSE projection head          & (564, 1)   & (564, 1536) & 867,840   & (1526, 1)   & (1536, 1)   & 2,345,472  \\
CLIP projection head         & (564, 1)   & (564, 1536) & 867,840   & (1526, 1)   & (1536, 1)   & 2,345,472  \\ \midrule
Total                        &            &             & 4,219,533 &             &             & 20,799,113 \\ \bottomrule
\end{tabular}
\end{table}

\begin{table}[]
\centering
\caption{Description of the brain module architectures used with MEG data.}
\label{tab:meg-architecture}
\begin{tabular}{@{}lllrllr@{}}
\toprule
                             & \multicolumn{3}{c}{\textbf{Medium}}  & \multicolumn{3}{c}{\textbf{Large}}     \\ \midrule
\multicolumn{1}{c}{\textbf{Layer}} &
  \multicolumn{1}{c}{\textbf{Input}} &
  \multicolumn{1}{c}{\textbf{Output}} &
  \multicolumn{1}{c}{\textbf{\# params}} &
  \multicolumn{1}{c}{\textbf{Input}} &
  \multicolumn{1}{c}{\textbf{Output}} &
  \multicolumn{1}{c}{\textbf{\# params}} \\ \midrule
Spatial attention            & (272, 180)   & (270, 180) & 552,960   & (272, 180)    & (270, 180)  & 552,960    \\
Linear projection            & (270, 180) & (50, 180)  & 13,550    & (270, 180)  & (396, 180)  & 107,316    \\
Subject layer                & (50, 180)  & (50, 180)  & 2,500     & (396, 180)  & (396, 180)  & 627,264    \\
Residual dilated conv blocks & (50, 180)  & (50, 180)  & 60,800    & (396, 180)  & (396, 180)  & 7,539,840  \\
1x1 conv block               & (50, 180)  & (152, 180) & 20,452    & (396, 180)  & (1411, 180) & 1,433,347  \\
Temporal aggregation         & (152, 180) & (152, 1)   & 181       & (1411, 180) & (1411, 1)   & 181        \\
MSE projection head          & (152, 1)   & (1536, 1)  & 235,008   & (1411, 1)   & (1536, 1)   & 2,168,832  \\
CLIP projection head         & (152, 1)   & (1536, 1)  & 235,008   & (1411, 1)   & (1536, 1)   & 2,168,832  \\ \midrule
Total                        &            &            & 1,120,459 &             &             & 14,598,572 \\ \bottomrule
\end{tabular}
\end{table}

\begin{table}[]
\centering
\caption{Description of the brain module architectures used with fMRI data.}
\label{tab:fmri-architecture}
\begin{tabular}{@{}lllrllr@{}}
\toprule
                     & \multicolumn{3}{c}{\textbf{Medium}}            & \multicolumn{3}{c}{\textbf{Large}}             \\ \midrule
\multicolumn{1}{c}{\textbf{Layer}} &
  \multicolumn{1}{c}{\textbf{Input}} &
  \multicolumn{1}{c}{\textbf{Output}} &
  \multicolumn{1}{c}{\textbf{\# params}} &
  \multicolumn{1}{c}{\textbf{Input}} &
  \multicolumn{1}{c}{\textbf{Output}} &
  \multicolumn{1}{c}{\textbf{\# params}} \\ \midrule
Subject layer        & (20484, 5) & (553, 5)  & 33,982,956            & (20484, 5) & (1552, 5) & 127,164,672           \\
TR layer             & (553, 5)   & (553, 5)  & 1,532,916             & (1552, 5)  & (1552, 5) & 12,054,384            \\
Residual conv blocks & (553, 5)   & (553, 5)  & 614,936               & -          & -         & \multicolumn{1}{l}{-} \\
Temporal aggregation & (553, 5)   & (553, 1)  & 6                     & (1552, 5)  & (1552, 1) & 6                     \\
Linear projection    & (553, 1)   & (1536, 1) & 850,944               & (1552, 1)  & (1536, 1) & 2,385,408             \\
MSE projection head  & (1536, 1)  & (1536, 1) & 2,360,832             & (1536, 1)  & (1536, 1) & 2,360,832             \\
CLIP projection head & (1536, 1)  & (1536, 1) & \multicolumn{1}{l}{-} & (1536, 1)  & (1536, 1) & 2,363,904             \\ \midrule
Total                &            &           & 39,342,590            &            &           & 146,329,206           \\ \bottomrule
\end{tabular}
\end{table}

\section{Reproduction of main results in the \emph{matched-trials} setting}
\label{sec:reprod_matched_trials}

\begin{figure}
    \centering
    \includegraphics[width=1.0\linewidth]{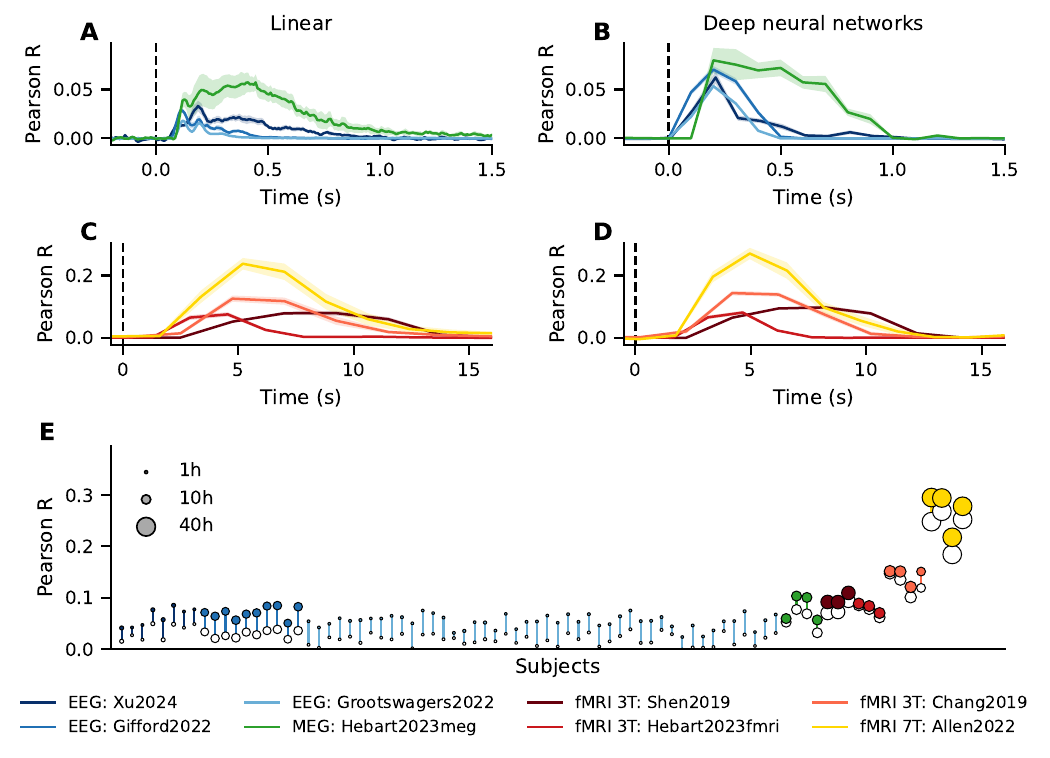}
    \caption{Stepwise image decoding analyses in the \emph{matched-trials} setting for (\textbf{left}) linear and (\textbf{right}) deep learning models.
    See description of \Cref{fig:stepwise_decoding}.
    }
\label{fig:stepwise_decoding_matched_trials}
\end{figure}

\begin{figure}
    \centering
    \includegraphics[width=\linewidth]{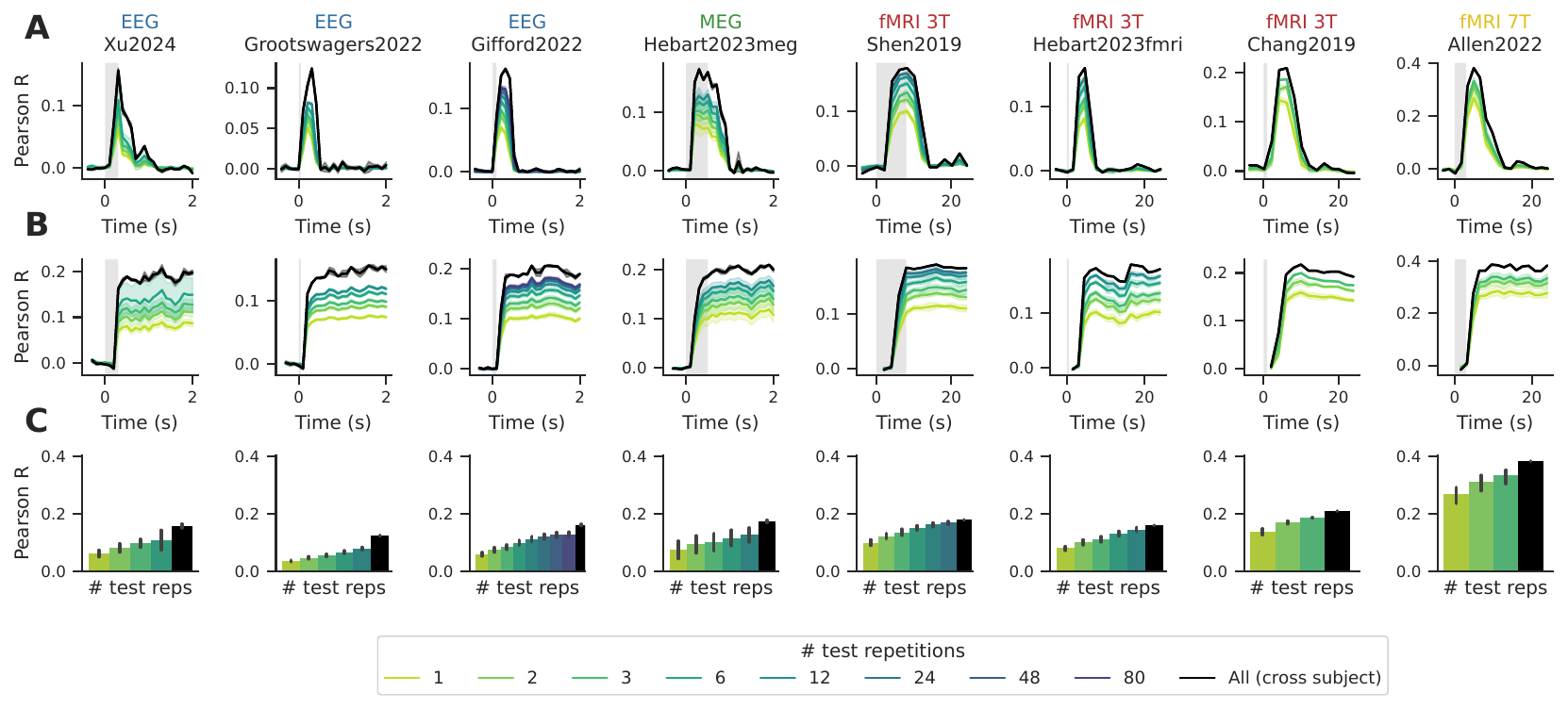}
    \caption{Decoding of the image embedding as a function of time (x-axis) and number of test-time repetitions (color) using deep learning models, in the \emph{matched-trials} setting.
    See description of \Cref{fig:test_trial_averaging}.
    }
    \label{fig:test_trial_averaging_matched_trials}
\end{figure}

We reproduce the results presented in the main text by using the \emph{matched-trials} setting (see \Cref{sec:matched_trials}), \ie we match the available quantity of data across datasets as closely as possible to allow comparing devices more directly.
\Cref{fig:stepwise_decoding_matched_trials} presents stepwise decoding results.
\Cref{fig:test_trial_averaging_matched_trials} presents results from the test-time averaging analyses.

\begin{figure}
    \centering
    \includegraphics[width=\linewidth]{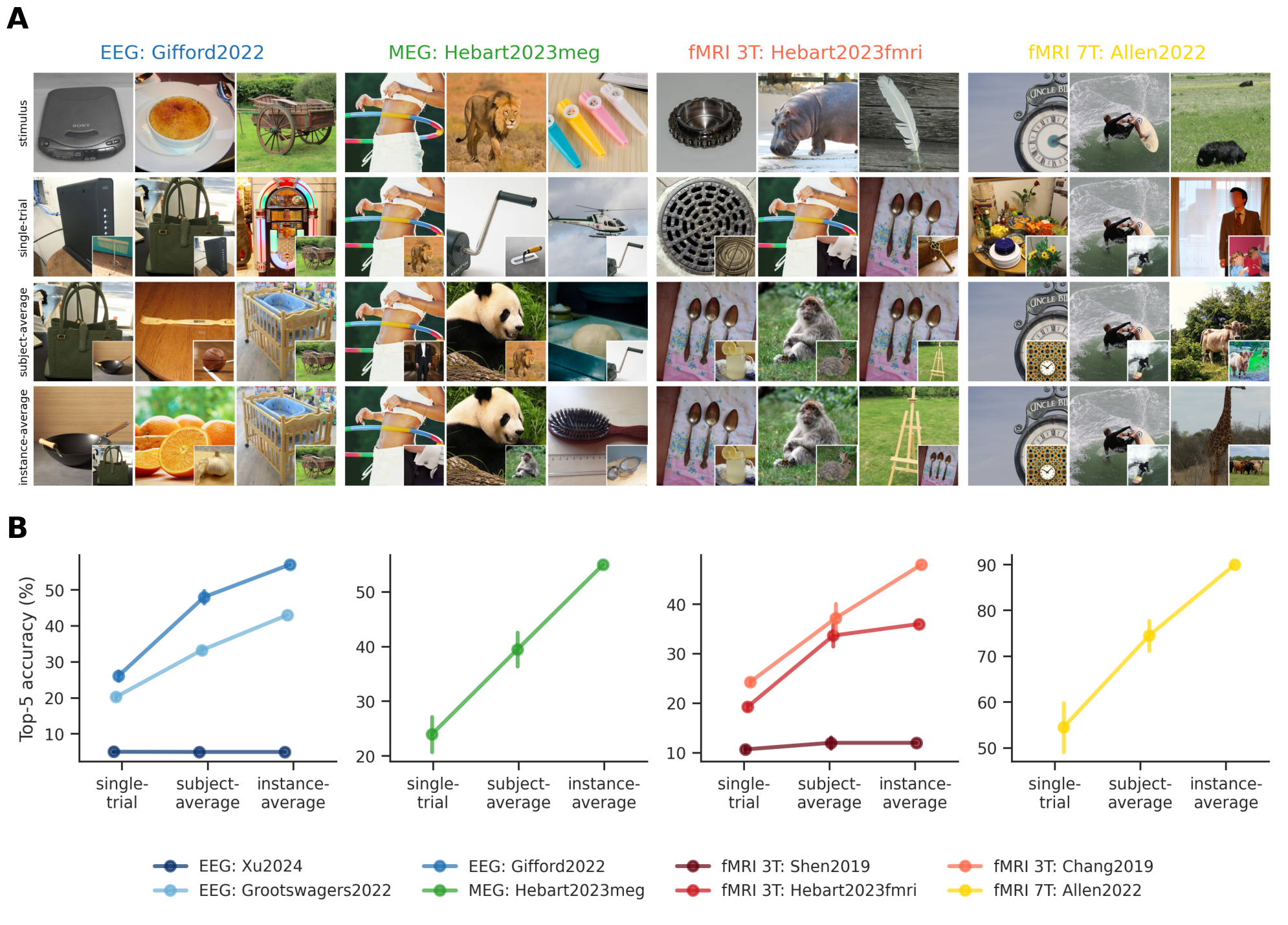}
    \caption{Image retrieval across devices in the \emph{matched-trials} setting. See description of \Cref{fig:best_retrievals}.
    }
    \label{fig:best_retrievals_matched}
\end{figure}

\begin{figure}
    \centering
    \includegraphics[width=\linewidth]{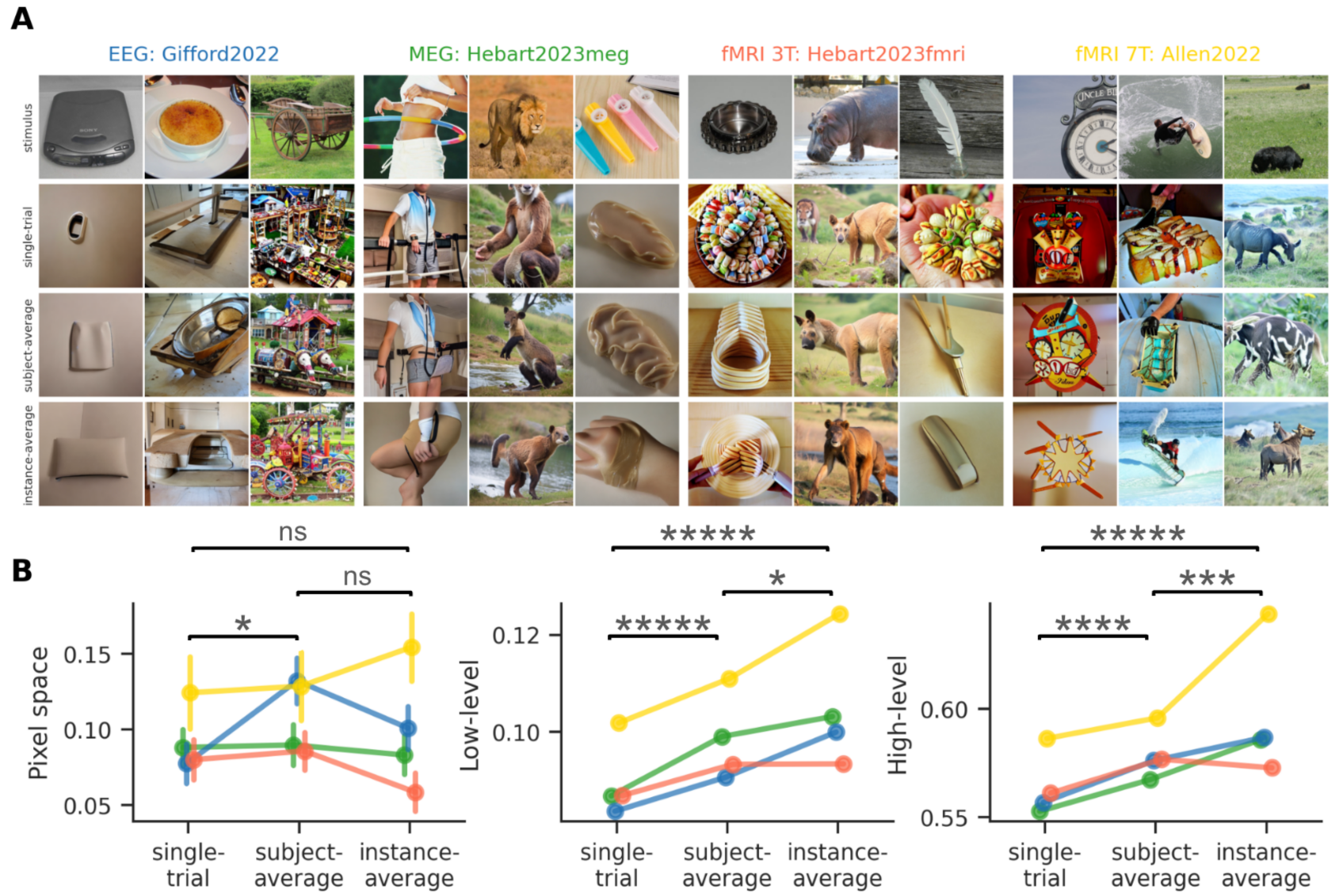}
    \caption{Image reconstruction across devices in the \emph{matched-trials} setting. See description of \Cref{fig:best_generations}.}
    \label{fig:best_generations_matched}
\end{figure}

\Cref{fig:best_retrievals_matched} and \ref{fig:best_generations_matched} show the retrieval and generation results.

\section{Cost estimation of neuroimaging data collection}
\label{app:cost}

We surveyed publicly available hourly cost estimates from research institutions or third-party providers that offer a neuroimaging data collection service.
For EEG, we used commercial third-party pricing for US-based services\footnote{\url{https://brainarcevaluations.com/pricing/} and \url{https://sadarpsych.com/services/data-collection/}.}.
For MEG, we used cost estimates from a European-based research center\footnote{\url{https://www.bcbl.eu/en/infrastructure-equipment/meg}.}, which we converted from EUR to USD at a rate of 1.1 (November 2024).
For fMRI 3T and 7T, we used cost estimates from a US-based research hospital center\footnote{\url{https://www.brighamandwomens.org/radiology/research-imaging-core/pricing}.}.
Finally, when a price range was provided (\eg corresponding to different pricing tiers for internal vs. external collaborators), we used the average between lowest and highest price for each device type.
Based on this information, hourly cost (in USD) was estimated at \$263
for EEG, \$550 for MEG, \$935 for 3T fMRI and \$1093 for 7T fMRI.

\section{Quality of image reconstructions across devices and test-time averaging strategies}

\begin{figure}
    \centering
    \includegraphics[width=\linewidth]{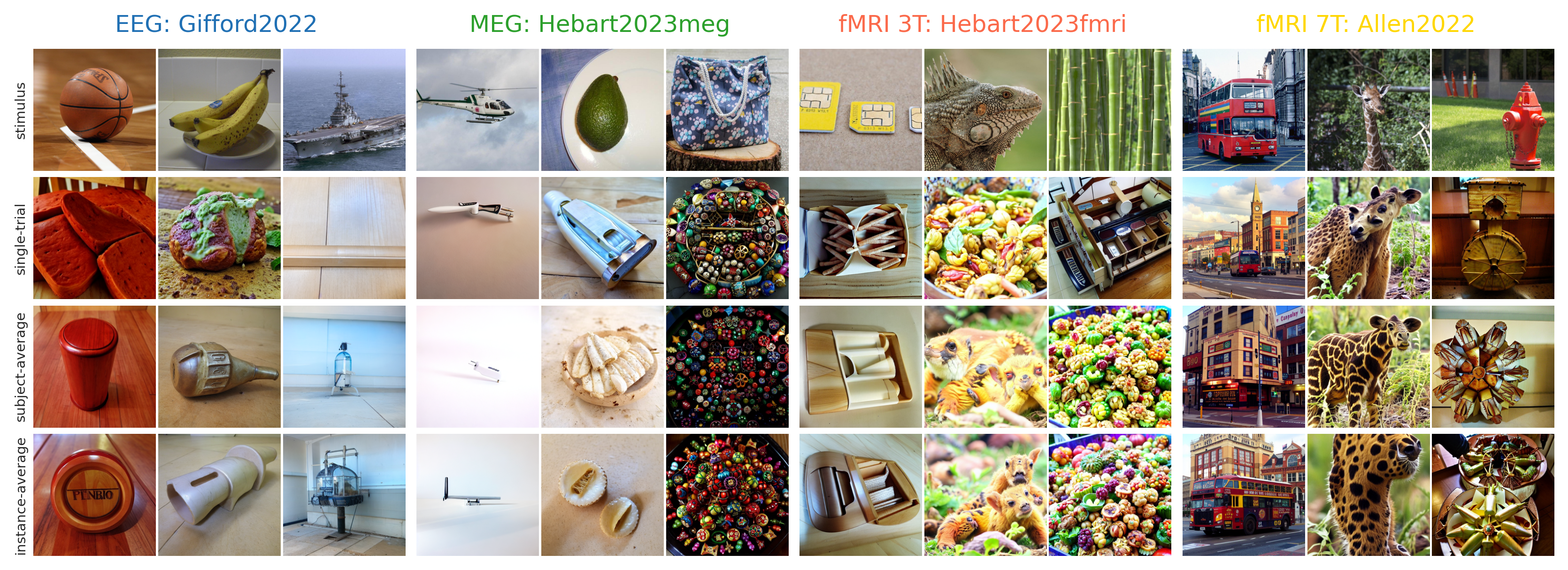}
    \caption{Image reconstruction across devices in the \emph{full-trials} setting. For each study, a sample of 3 stimuli showing various qualities of reconstructions obtained (left: good quality, middle: lower quality, right: failure case) from single-trial brain signals, and two increasingly large aggregations (averaging embedding predictions at subject-level and at instance-level).}
    \label{fig:good_med_bad_generations_matched}
\end{figure}

\begin{figure}
    \centering
    \includegraphics[width=\linewidth]{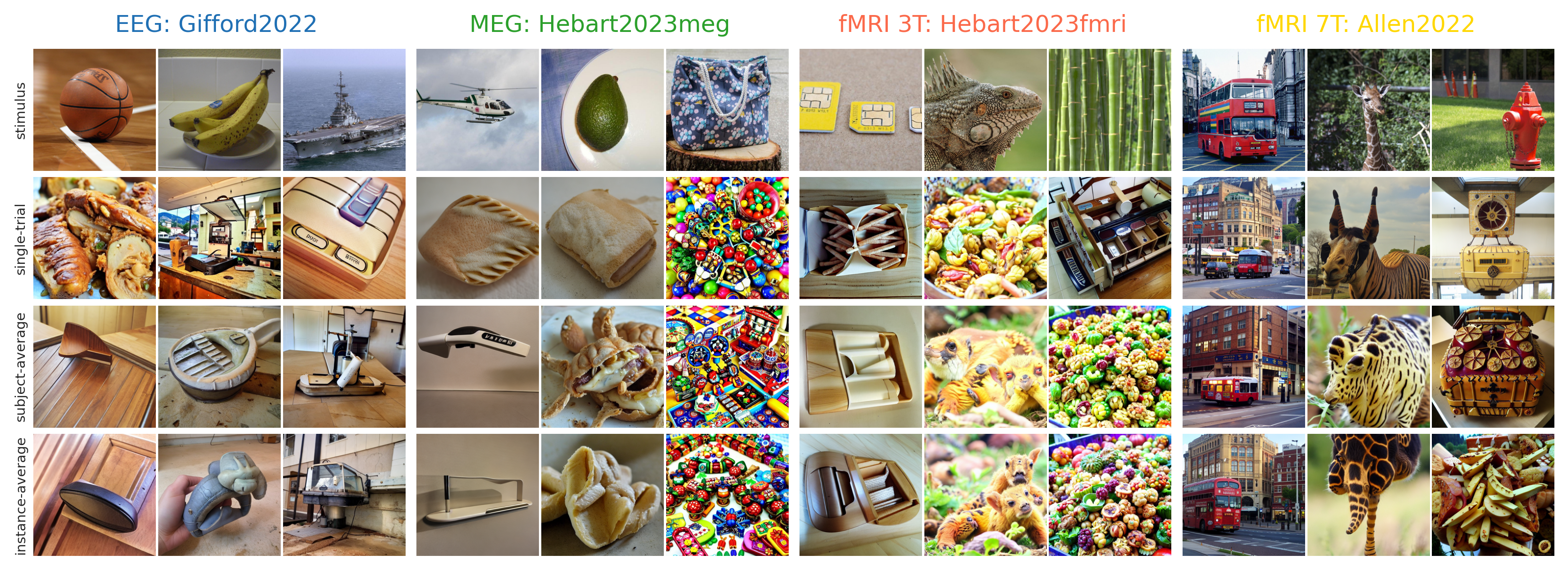}
     \caption{Image reconstruction across devices in the \emph{matched-trials} setting. For each study, a sample of 3 stimuli showing various qualities of reconstructions obtained (left: good quality, middle: lower quality, right: failure case) from single-trial brain signals, and two increasingly large aggregations (averaging embedding predictions at subject-level and at instance-level).}
    \label{fig:good_med_bad_generations}
\end{figure}

\Cref{fig:good_med_bad_generations} shows for each device the reconstructions we obtain for a sample of three stimuli. These three stimuli were chosen to present a broader view of reconstruction quality, compared to the best-case cherry-picking of \Cref{fig:best_generations}. \Cref{fig:good_med_bad_generations_matched} shows the same results, but in the \emph{matched-trials} data configuration.
These results confirm that aggregating predictions benefits high- and medium-quality reconstructions, though it is unclear whether it actually benefits bad reconstructions.

Finally, \Cref{tab:generation_metrics} and \ref{tab:generation_metrics_matched} presents the generation metrics defined in \Cref{subsec:image_reconstruction_methods} and the additional metrics reported in recent works \citep{ozcelik2023brain,scotti2023reconstructing,scotti2024mindeye2} for each representative dataset, each test-time averaging strategy and for each of the \emph{all-trials} and \emph{matched-trials} settings.

\begin{table}[ht]
\centering
\caption{Quantitative evaluation of reconstruction quality from each reconstructed study with different test-time averaging strategies, in the \emph{all-trials} setting (see \Cref{subsec:image_reconstruction_methods} for AlexNet-2-R and CLIP-Final-R metric definitions).}
\label{tab:generation_metrics}
\resizebox{\textwidth}{!}{%
\begin{tabular}{@{}lllllllllll@{}}
\toprule
Study            & Averaging        & PixCorr $\uparrow$ & SSIM $\uparrow$  & AlexNet-2 $\uparrow$ & AlexNet-5 $\uparrow$ & CLIP-Final $\uparrow$ & InceptionV3 $\uparrow$ & SwAV $\downarrow$ & AlexNet-2-R $\uparrow$ & CLIP-Final-R $\uparrow$ \\ \midrule
Gifford2022      & single-trial     & 0.084   & 0.252 & 0.681     & 0.722     & 0.621      & 0.561               & 0.631        & 0.096         & 0.567     \\
                 & subject-average  & 0.138   & 0.265 & 0.793     & 0.86      & 0.754      & 0.695               & 0.573        & 0.118         & 0.596     \\
                 & instance-average & 0.126   & 0.249 & 0.822     & 0.883     & 0.777      & 0.728               & 0.571        & 0.125         & 0.608     \\
\midrule
Hebart2023meg      & single-trial     & 0.064   & 0.292 & 0.702     & 0.762     & 0.696      & 0.601               & 0.61         & 0.098         & 0.565     \\
                 & subject-average  & 0.081   & 0.294 & 0.799     & 0.869     & 0.743      & 0.685               & 0.573        & 0.12          & 0.591     \\
                 & instance-average & 0.083   & 0.286 & 0.78      & 0.892     & 0.807      & 0.701               & 0.569        & 0.116         & 0.603     \\
\midrule
Hebart2023fmri   & single-trial     & 0.08    & 0.234 & 0.62      & 0.667     & 0.682      & 0.596               & 0.642        & 0.087         & 0.561     \\
                 & subject-average  & 0.086   & 0.25  & 0.658     & 0.741     & 0.718      & 0.608               & 0.622        & 0.093         & 0.577     \\
                 & instance-average & 0.058   & 0.233 & 0.67      & 0.754     & 0.736      & 0.62                & 0.62         & 0.093         & 0.573     \\
\midrule
Allen2022    & single-trial     & 0.119   & 0.19  & 0.742     & 0.793     & 0.781      & 0.768               & 0.533        & 0.11          & 0.588     \\
                 & subject-average  & 0.11    & 0.19  & 0.75      & 0.807     & 0.839      & 0.793               & 0.503        & 0.118         & 0.616     \\
                 & instance-average & 0.192   & 0.221 & 0.842     & 0.89      & 0.888      & 0.846               & 0.452        & 0.139         & 0.661     \\ \bottomrule
\end{tabular}
}
\end{table}

\begin{table}[ht]
\centering
\caption{Quantitative evaluation of reconstruction quality from each reconstructed study with different test-time averaging strategies, in the \emph{matched-trials} setting.}
\label{tab:generation_metrics_matched}
\resizebox{\textwidth}{!}{%
\begin{tabular}{@{}lllllllllll@{}}
\toprule
Study            & Averaging        & PixCorr $\uparrow$ & SSIM $\uparrow$  & AlexNet-2 $\uparrow$ & AlexNet-5 $\uparrow$ & CLIP-Final $\uparrow$ & InceptionV3 $\uparrow$ & SwAV $\downarrow$ & AlexNet-2-R $\uparrow$ & CLIP-Final-R $\uparrow$ \\ \midrule
Gifford2022      & single-trial     & 0.077   & 0.22  & 0.621     & 0.687     & 0.625      & 0.597               & 0.648        & 0.084         & 0.557     \\
                 & subject-average  & 0.132   & 0.246 & 0.678     & 0.734     & 0.671      & 0.63                & 0.624        & 0.091         & 0.576     \\
                 & instance-average & 0.101   & 0.239 & 0.711     & 0.776     & 0.716      & 0.652               & 0.61         & 0.1           & 0.587     \\
\midrule
Hebart2023meg      & single-trial     & 0.088   & 0.258 & 0.636     & 0.729     & 0.64       & 0.589               & 0.642        & 0.087         & 0.553     \\
                 & subject-average  & 0.09    & 0.259 & 0.695     & 0.772     & 0.715      & 0.631               & 0.621        & 0.099         & 0.567     \\
                 & instance-average & 0.083   & 0.264 & 0.733     & 0.823     & 0.751      & 0.593               & 0.606        & 0.103         & 0.586     \\
\midrule
Hebart2023fmri   & single-trial     & 0.08    & 0.234 & 0.62      & 0.667     & 0.682      & 0.596               & 0.642        & 0.087         & 0.561     \\
                 & subject-average  & 0.086   & 0.25  & 0.658     & 0.741     & 0.718      & 0.608               & 0.622        & 0.093         & 0.577     \\
                 & instance-average & 0.058   & 0.233 & 0.67      & 0.754     & 0.736      & 0.62                & 0.62         & 0.093         & 0.573     \\
\midrule
Allen2022    & single-trial     & 0.124   & 0.185 & 0.672     & 0.742     & 0.753      & 0.708               & 0.551        & 0.102         & 0.586     \\
                 & subject-average  & 0.129   & 0.19  & 0.728     & 0.799     & 0.78       & 0.772               & 0.535        & 0.111         & 0.596     \\
                 & instance-average & 0.154   & 0.22  & 0.773     & 0.858     & 0.862      & 0.837               & 0.471        & 0.124         & 0.644     \\ \bottomrule
\end{tabular}
}
\end{table}

\end{document}